\documentclass{jfm}

\usepackage{graphicx}
\usepackage{newtxtext}
\usepackage{newtxmath}
\usepackage{natbib}
\usepackage{hyperref}
\usepackage{soul}
\hypersetup{
    colorlinks = true,
    urlcolor   = blue,
    citecolor  = black,
}

\newcommand{\RomanNumeralCaps}[1]
\linenumbers

\title{Asymptotic Nusselt numbers for internal flow in the Cassie state}

\author{Daniel Kane\aff{1},
  Marc Hodes\aff{1},
  Martin Z. Bazant\aff{2,3}
 \and Toby L. Kirk\aff{4}\corresp{\email{tlk12@ic.ac.uk}}}

\affiliation{\aff{1}Department of Mechanical Engineering, Tufts University, Medford, MA, US
\aff{2}Department of Chemical Engineering, Massachusetts Institute of Technology, Cambridge, MA, US
\aff{3}Department of Mathematics, Massachusetts Institute of Technology, Cambridge, MA, US
\aff{4}Department of Mathematics, Imperial College London, London SW7 2AZ, UK
}

\begin{document}
\maketitle

\begin{abstract}
We consider laminar, fully-developed, Poiseuille flows of liquid in the Cassie state through diabatic, parallel-plate microchannels symmetrically textured with isoflux ridges. Through the use of matched asymptotic expansions we analytically develop expressions for (apparent hydrodynamic) slip lengths and variously-defined Nusselt numbers. Our small parameter ($\epsilon$) is the pitch of the ridges divided by the height of the microchannel. When the ridges are oriented parallel to the flow, we quantify the error in the Nusselt number expressions in the literature and provide a new closed-form result. The latter is accurate to $O\left(\epsilon^2\right)$ and valid for any solid (ridge) fraction, whereas those in the current literature are accurate to $O\left(\epsilon^1\right)$ and 
breakdown in the important limit when solid fraction approaches zero. When the ridges are oriented transverse to the (periodically fully-developed) flow, the error associated with neglecting inertial effects in the slip length is shown to be $O\left(\epsilon^3\mathrm{Re}\right)$, where 
$\mathrm{Re}$ is the channel-scale Reynolds number based on its hydraulic diameter. The corresponding  Nusselt number expressions are new and their accuracy is shown to be dependent on 
Reynolds number, P{\'e}clet number and Prandtl number in addition to $\epsilon$.
Manipulating the solution to the inner temperature problem encountered in the vicinity of the ridges shows that classic results for thermal spreading resistance are better expressed in terms of polylogarithm functions.
\end{abstract}


{\bf MSC Codes }  {\it(Optional)} Please enter your MSC Codes here

\section{Introduction}
\subsection{Background and Motivation}
Internal flows through microchannels textured with hydrophobic ridges, pillars, etc.~to maintain liquid in the Cassie state for lubrication have received considerable attention in the recent 
literature \citep{Rothstein-10,Lee-16}. The majority of studies have considered adiabatic flows, but studies of diabatic ones, discussed at length by \citet{Game-18}, are also of interest. Applications include liquid cooling of microelectronics, where, beneficially, lubrication decreases caloric resistance and, detrimentally, reduced solid-liquid area for heat transfer increases convective resistance.  (It's noteworthy that
transition from laminar to turbulent flow has the opposite effect, i.e., it increases caloric resistance and decreases convective resistance.) Notably, \citet{Lam-15} showed that a net enhancement is likely via judicious texturing of microchannel with a suherphydrophobic (SH) surface when the coolant is a liquid metal
and thus the caloric resistance is dominant.  The key engineering parameters in
such problems are the Poiseuille number (from which caloric resistance follows) and variously-defined Nusselt numbers 
(from which convective resistances follow). This study provides more convenient and accurate closed-form expressions for these numbers
than available in the literature. We note that, to be consistent with the literature, we provide the
(dimensionless, apparent hydrodynamic) slip lengths 
rather than the Poiseuille number itself; see, e.g., \citet{Enright-14} for conversion between the two parameters.

\subsection{Problems}
We consider laminar, Poiseuille flow of a liquid in the Cassie state through a parallel-plate microchannel symmetrically textured with equally-spaced, isoflux, no-slip ridges aligned parallel or transverse to the flow direction.  The flow is hydrodynamically- and thermally-fully-developed for parallel ridges.  The
flow and temperature problems are periodically fully-developed for transverse ridges. We assume constant thermophysical properties and ignore viscous dissipation. We further assume that menisci are flat, shear-free and adiabatic and we ignore the effects of thermocapillary stress along them. By implication, phase change effects along menisci are neglected. A brief discussion of thermocapillary and phase change effects along curved menisci is provided in our Conclusions. 

\subsection{Scope}
The assumption that multi-dimensional effects due to the ridges on the velocity and temperature fields are confined to an ``inner region'' near them has been the basis for many analyses of internal liquid flows in the Cassie state  when the microchannel height is large compared to the ridge pitch. Conveniently, this allows the use of solutions for semi-infinite domains for the inner region because it's small compared to the ``outer region,'' where the velocity field becomes, asymptotically, unidirectional and one dimensional and the temperature field is two-dimensional. Correspondingly, the flow is  governed by Laplace's equation and the Stokes equations in the inner region for parallel and transverse ridges, respectively, and by the same one-dimensional Poisson equation in both outer ones.  In turn,
asymptotically, the temperature field (heat transfer) in both cases is governed by Laplace's equation 
in the inner region and by the same two-term advection-diffusion equation in the outer one. We formally quantify the validity of these assumptions through the use of matched asymptotic expansions. Specifically, only assuming that the ridge pitch, $2d$, is small compared to the microchannel height, $2H$, we scale
the effects of relevant  hydrodynamic and thermal phenomena in the two regions  with respect to the small parameter $\epsilon=d/H$
and, in the case of transverse ridges, Reynolds number (Re) and P{\'e}clet number (Pe). 
Significantly, unlike in most studies, we  provide the error terms for our slip length and Nusselt number results.
Also, we provide a new closed-form (mean) Nusselt number expression  accurate to $O\left(\epsilon^2\right)$ for parallel ridges and one with its accuracy dependent upon 
Reynolds number, P{\'e}clet number and Prandtl number, in addition to $\epsilon$,
for transverse ridges. Both of these results remain valid as solid fraction approaches zero,
a significant advancement. We also show that a classic spreading resistance result which
resolves our inner thermal problem can be better expressed in terms of polylogarithm (special) functions.  All of our results are compared against those in the literature.

\section{Parallel Ridges}
\label{sec:parallel}
\begin{figure}
\begin{centering}
\includegraphics[width=14cm]{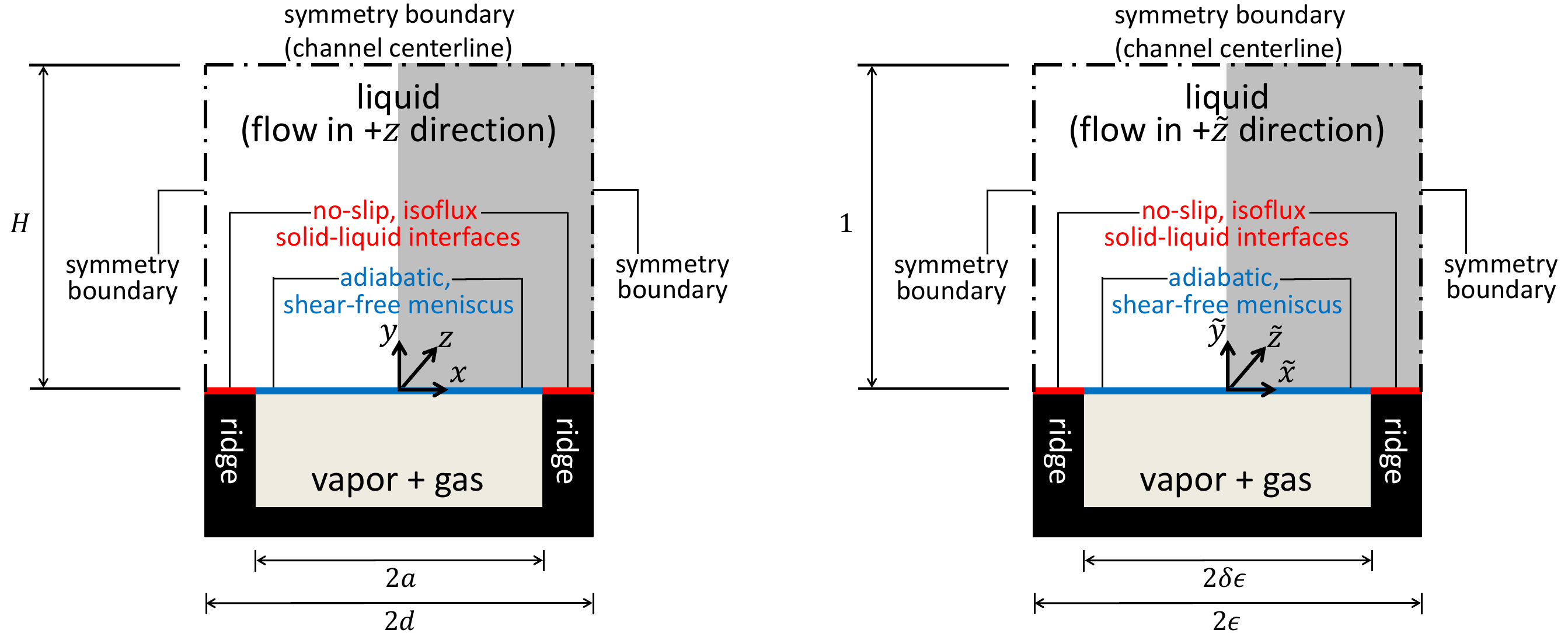}
\par\end{centering}
\protect\caption{Schematic of the dimensional geometry of the parallel-ridge problem (left) and 
the corresponding dimensionless geometry (right). The (planar) domain is shown in gray. A (constant) streamwise-pressure
gradient is imposed along the $z$-direction.\label{fig:Parallel_Domain-and-boundary}}
\end{figure}

A schematic of the (dimensional) geometry of the parallel-ridge problem is shown in Fig.~\ref{fig:Parallel_Domain-and-boundary} (left). The unidirectional and fully-developed flow in the $z$-direction is driven by a prescribed pressure gradient, $\mathrm{d}p/\mathrm{d}z$, a negative constant. The problem is symmetric in the 
$x$-direction and the width of the meniscus is $2a$ and that of the ridges is $2\left(d-a\right)$. Our domain extends from $x = 0$ to $x = d$ and from $y=0$ to $y=H$, the channel centerline, on account of symmetries. Vapor and/or noncondensable gas may be present in the cavity between the ridges, but we ignore it.

\subsection{Hydrodynamic Problem}
\label{sub:Hydrodynamic}
The dimensional form of the streamwise-momentum equation is
\begin{equation}
\frac{\partial^{2}w}{\partial x^{2}}+\frac{\partial^{2}w}{\partial y^{2}}=\frac{1}{\mu}\frac{\mathrm{d} p}{\mathrm{d} z},
\end{equation}
where $w$ is the streamwise velocity and $\mu$ is the dynamic viscosity of the liquid. The shear-free and no-slip boundary conditions along the meniscus and solid-liquid interface, respectively, and symmetry ones elsewhere on the domain manifest themselves as
\begin{align}
\frac{\partial w}{\partial y}&=0\,\,\,\mathrm{at}\,\,\, y=0\,\,\,\mathrm{for}\,\,\,0<x<a\\
w&=0\,\,\,\mathrm{at}\,\,\, y=0\,\,\,\mathrm{for}\,\,\, a<x<d\\
\frac{\partial w}{\partial y}&=0\,\,\,\mathrm{at}\,\,\, y=H\,\,\,\mathrm{for}\,\,\,0<x<d\\
\frac{\partial w}{\partial x}&=0\,\,\,\mathrm{at}\,\,\, x=0\,\,\,\mathrm{and}\,\,\, x=d\,\,\,\mathrm{for}\,\,\,0<y<H,
\end{align}
respectively.
We nondimensionalize lengths by $H$ and indicate that a quantity is nondimensional by placing a tilde over it. The dimensionless geometry of the problem is shown
in Fig.~\ref{fig:Parallel_Domain-and-boundary} (right). Moreover, we nondimensionalize $w$ by $\left(-\mathrm{d}p/\mathrm{d}z\right)H^2/\left(2\mu\right)$.  Then, the hydrodynamic problem becomes 
\begin{equation}
\frac{\partial^{2}\tilde{w}}{\partial\tilde{x}^{2}}+\frac{\partial^{2}\tilde{w}}{\partial\tilde{y}^{2}}=-2
\end{equation}
subject to
\begin{align}
\frac{\partial\tilde{w}}{\partial\tilde{y}}&=0\,\,\,\mathrm{at}\,\,\,\tilde{y}=0\,\,\,\mathrm{for}\,\,\,0 < \tilde{x}<\delta\epsilon \\
\tilde{w}&=0\,\,\,\mathrm{at}\,\,\,\tilde{y}=0\,\,\,\mathrm{for}\,\,\,\delta\epsilon<\tilde{x}<\epsilon \\
\frac{\partial\tilde{w}}{\partial\tilde{y}}&=0\,\,\,\mathrm{at}\,\,\,\tilde{y}=1\,\,\,\mathrm{for}\,\,\,0<\tilde{x}<\epsilon \\
\frac{\partial\tilde{w}}{\partial\tilde{x}}&=0\,\,\,\mathrm{at}\,\,\,\tilde{x}=0\,\,\,\mathrm{and}\,\,\,\tilde{x}=\epsilon\,\,\,\mathrm{for}\,\,\,0<\tilde{y}<1,
\end{align}
where $\delta=a/d$ is the cavity fraction. 

\begin{figure}
\begin{centering}
\includegraphics[width=10cm]{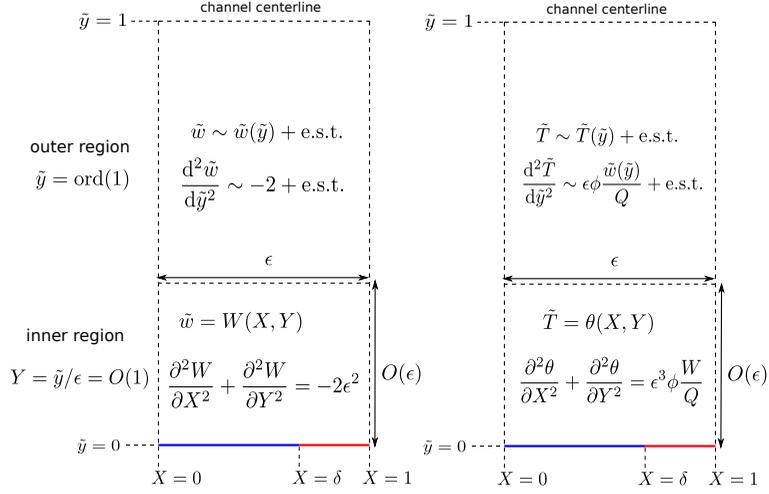}
\par\end{centering}
\caption{Depiction of matched asymptotic expansions for flow (left) and temperature (right)
problems.}
\label{fig:mae}
\end{figure}

We resolve the velocity field using a matched
asymptotic expansion for $\epsilon \ll 1$ in reference to Fig.~\ref{fig:mae} (left). 
The outer region occupies the majority of the domain and it
will be shown that, asymptotically, the velocity field there is one-dimensional
and parabolic. The spatially-periodic portion of the velocity field on account of the ridges 
is shown to be confined to an inner region in the domain where $\tilde{y}$
is of comparable scale to the ridge pitch. 

Our analysis follows that of \citet{Hodes-17}, but both sides rather than one side of the microchannel are textured and thermocapillary stresses along menisci are absent. The extents of the rectangular domain are made independent of the small parameter by working in terms of $X = \tilde{x}/\epsilon$ such that the problem becomes 
\begin{equation}
\frac{1}{\epsilon^{2}}\frac{\partial^{2}\tilde{w}}{\partial X^{2}}+\frac{\partial^{2}\tilde{w}}{\partial\tilde{y}^{2}}=-2\label{eq:Parallel outer momentum}
\end{equation}
subject to
\begin{align}
\frac{\partial\tilde{w}}{\partial\tilde{y}}&=0\,\,\,\mathrm{at}\,\,\,\tilde{y}=0\,\,\,\mathrm{for}\,\,\,0<X<\delta \\
\tilde{w}&=0\,\,\,\mathrm{at}\,\,\,\tilde{y}=0\,\,\,\mathrm{for}\,\,\,\delta<X<1 \\
\frac{\partial\tilde{w}}{\partial\tilde{y}}&=0\,\,\,\mathrm{at}\,\,\,\tilde{y}=1\,\,\,\mathrm{for}\,\,\,0<X<1 \label{eq:centsym}\\
\frac{\partial\tilde{w}}{\partial X}&=0\,\,\,\mathrm{at}\,\,\, X=0\,\,\,\mathrm{and}\,\,\, X=1\,\,\,\mathrm{for}\,\,\,0<\tilde{y}<1.
\end{align}

\subsubsection{Outer Region\label{sub:Parallel-Hydro-Outer-Region}}
The outer region of the domain is where $X=O(1),\,\tilde{y}=\mathrm{ord}(1)$, i.e., strictly of order unity as $\epsilon\rightarrow0$. It's sufficiently far from the mixed boundary condition imposed at $\tilde{y}=0$ that the velocity field is one-dimensional, i.e., $\tilde{w} = \tilde{w}\left(\tilde{y}\right)$, as justified henceforth. 
Equation\,(\ref{eq:Parallel outer momentum})
reduces, to leading order, to $\partial^{2}\tilde{w}/\partial X^{2}=0$.
Integrating, satisfying the symmetry boundary condition in $X$ and integrating again shows
that, to leading order, $\tilde{w}=\tilde{w}\left(\tilde{y}\right)$. Thus, the $\partial^2\tilde{w}/\partial\tilde{y}^2$ and $-2$ terms in  (\ref{eq:Parallel outer momentum}) balance, implying that $\tilde{w}=O(1)$. We thus, as an ansatz, expand the velocity as a regular power series in $\epsilon$ as
per
\begin{equation}
\tilde{w}=\sum_{n=0}^{\infty}\epsilon^{n}\tilde{w}_{n}+ \mathrm{e.s.t.},~~~\epsilon \rightarrow 0,\label{eq:Outer velo expand}
\end{equation}
where $\tilde{w}_{n}=O(1)$ for all $n\geq0$ and e.s.t.~denotes exponentially small terms that are smaller than any algebraic order of $\epsilon$. We do n0t attempt to solve for these terms, and only include them for error quantification. Substituting Eq.\,(\ref{eq:Outer velo expand})
into Eq.\,(\ref{eq:Parallel outer momentum}), yields an equation at each (algebraic) order in $\epsilon$ as per
\begin{align}
O\left(\epsilon^{-2}\right):~~~~~~~~~~~~&\frac{\partial^{2} \tilde{w}_{0}}{\partial X^2}=0 \label{eq:order0} \\
O\left(\epsilon^{-1}\right):~~~~~~~~~~~~&\frac{\partial^{2} \tilde{w}_{1}}{\partial X^2}=0 \\ 
O\left(\epsilon^{0}\right):~~~~~~~~~~~~&\frac{\partial^{2} \tilde{w}_{2}}{\partial X^2}+\frac{\partial^{2} \tilde{w}_{0}}{\partial \tilde{y}^{2}}=-2 \label{eq:order0a}\\
O\left(\epsilon^{n}\right):~~~~~~~~~~~~&\frac{\partial^{2} \tilde{w}_{n+2}}{\partial X^2}+\frac{\partial^{2} \tilde{w}_{n}}{\partial \tilde{y}^{2}}=0,~~~~~~~n\geq 1.\label{eq:ordern}
\end{align}

The symmetry conditions at $X=0,1$ apply to all orders of $\tilde{w}$. Integrating the $O\left(\epsilon^{-1}\right)$ problem and applying them shows that $\tilde{w}_{1}$, like $\tilde{w}_{0}$, is purely a function of $\tilde{y}$. The $O\left(\epsilon^{0}\right)$
problem shows that $\partial^{2}\tilde{w}_{2}/\partial X^{2}$ is purely a function of $\tilde{y}$ and on account of the symmetry conditions so is $\tilde{w}_{2}$.
By induction in $n$, the remaining problems show that $\tilde{w}_{n}=\tilde{w}{}_{n}(\tilde{y})$
for $n\geq0$. Thus,  (\ref{eq:order0a})-(\ref{eq:ordern}) collapse to
\begin{align}
\frac{\mathrm{d}^{2}\tilde{w}_{0}}{\mathrm{d}\tilde{y}^{2}}&=-2
\\
\frac{\mathrm{d}^{2}\tilde{w}_{n}}{\mathrm{d}\tilde{y}^{2}}&=0,\,\,\,\,\,\, n\geq1.
\end{align}
Integrating and applying the symmetry condition at the channel centerline as per  (\ref{eq:centsym}) yields
\begin{equation}
\tilde{w}_{0}\left(\hat{y}\right)=-\tilde{y}^{2}+2\tilde{y}+C_{0},
\end{equation}
\begin{equation}
\tilde{w}_{n}\left(\tilde{y}\right)=C_{n},
\end{equation}
where $C_{n}$ are constants. Thus the solution to the outer problem is
\begin{equation}
\tilde{w}\sim -\tilde{y}^{2}+2\tilde{y}+C\left(\epsilon\right)+\mathrm{e.s.t.}, \label{eq:Outer_velo_parallel}
\end{equation}
where
\begin{equation}
C\left(\epsilon\right)=\sum_{n=0}^{\infty}C_{n}\epsilon^{n}.
\end{equation}
We turn to the inner problem to satisfy the mixed boundary condition at the composite interface, i.e.,
the meniscus and the solid-liquid interface.

\subsubsection{Inner Region\label{sub:Hydrodynamic-Inner-Region}}
The inner region of the domain is where $\tilde{x} \sim \tilde{y} =O(\epsilon)$, i.e., they're on the scale of the pitch of the ridges, as $\epsilon \rightarrow0$. This is sufficiently close   to the mixed boundary condition imposed at $\tilde{y}=0$ that the velocity field is two-dimensional, i.e., $\tilde{w} = \tilde{w}\left(\tilde{x},\tilde{y}\right)$. Our stretched coordinates are $X=\tilde{x}/\epsilon,\, Y=\tilde{y}/\epsilon$ such that the inner region corresponds to $X\sim Y=O(1)$ as $\epsilon\rightarrow0$
as per Fig.~\ref{fig:mae} (left).
Since the $X$ and $Y$ scales are of the same order, the viscous stress (Laplacian) terms are of the same order in the momentum equation. The velocity scale for the inner problem follows from the outer solution
as $\tilde{y}$ is decreased to $O\left(\epsilon\right)$. Substituting $\tilde{y}=\epsilon Y$
into  (\ref{eq:Outer_velo_parallel}) yields
\begin{equation}
\tilde{w}\sim-\epsilon^{2}Y^{2}+2\epsilon Y+C\left(\epsilon\right)+\mathrm{e.s.t.}\label{eq:outer_velo_parallel_inner_format}
\end{equation}
Hence, when $C(\epsilon)=O(1)$,
i.e., ~$C_{0}\neq0$ as required by  (\ref{eq:Parallel outer momentum}), $\tilde{w}=\mathrm{O(1)}$ in the inner region.  

Expressing the inner velocity field as $\tilde{w}=W(X,Y)$, the problem becomes
\begin{equation}
\frac{\partial^{2}W}{\partial X^{2}}+\frac{\partial^{2}W}{\partial Y^{2}}=-2\epsilon^{2}\label{eq:Parallel_inner_hydro_govern}
\end{equation}
subject to
\begin{align}
\frac{\partial W}{\partial Y}&=0\,\,\mathrm{at}\,\, Y=0\,\,\mathrm{for}\,\,0<X<\delta\\
W&=0\,\,\mathrm{at}\,\, Y=0\,\,\mathrm{for}\,\,\delta<X<1\\
W&\sim-\epsilon^{2}Y^{2}+2\epsilon Y+C(\epsilon)+\mathrm{e.s.t.}\,\,\mathrm{as}\,\, Y\rightarrow\infty\,\,\mathrm{for}\,\,0<X<1 \label{eq:Parallel_hydro_match}\\
\frac{\partial W}{\partial X}&=0\,\,\,\mathrm{at}\,\,\, X=0\,\,\,\mathrm{and}\,\,\, X=1\,\,\,\mathrm{for}\,\,\,Y>0,
\end{align}
where  (\ref{eq:Parallel_hydro_match}) is the (Van Dyke) matching condition in the overlap region, $\epsilon \ll \tilde{y} \ll 1$ (or $1 \ll Y \\ \epsilon^{-1})$.
Expressing $W$ as
\begin{equation}
W=-\epsilon^{2}Y^{2}+2\epsilon\widehat{W},\label{eq:Inner_velo_incomp}
\end{equation}
$\widehat{W}$ must satisfy 
\begin{equation}
\nabla^{2}\widehat{W}=0
\end{equation}
subject to
\begin{align}
\frac{\partial\widehat{W}}{\partial Y}&=0\,\,\mathrm{at}\,\, Y=0\,\,\mathrm{for}\,\,0<X<\delta\\
\widehat{W}&=0\,\,\mathrm{at}\,\, Y=0\,\,\mathrm{for}\,\,\delta<X<1\\
\widehat{W}&\sim Y+\frac{C\left(\epsilon\right)}{2\epsilon}+\mathrm{e.s.t.}\,\,\mathrm{as}\,\, Y\rightarrow\infty\\
\frac{\partial\widehat{W}}{\partial X}&=0\,\,\,\mathrm{at}\,\,\, X=0\,\,\,\mathrm{and}\,\,\, X=1\,\,\,\mathrm{for}\,\,\,0<Y<\infty.
\end{align}
This $\widehat{W}$ problem is the superposition of a one-dimensional linear-shear flow over a smooth surface and a perturbation to it which accommodates a mixed boundary condition and manifests itself with a constant mean velocity over the width of the domain. It has been solved using a conformal map by \citet{Philip-72,Philip-72b}. The solution up to the exponentially small error in the 
matching condition is
\begin{equation}
\widehat{W}=\mathrm{Im}\left\{ \frac{2}{\pi}\mathrm{cos}^{-1}\left[\frac{\mathrm{cos}\left(\pi \Theta_{\parallel}/2\right)}{\mathrm{cos}\left(\pi\delta/2\right)}\right]\right\}+\mathrm{e.s.t.}, 
\label{eq:What}
\end{equation}
where $\Theta_{\parallel}=X+\mathrm{i}Y$, and
\begin{equation}
\widehat{W}\sim Y+\lambda+\mathrm{e.s.t.}\,\,\mathrm{as}\,\, Y\rightarrow\infty,
\end{equation}
where
\begin{equation}
\lambda=\frac{2}{\pi}\mathrm{ln}\left[\mathrm{sec}\left(\delta\pi/2\right)\right].\label{eq:F definitions}
\end{equation}
It follows that 
\begin{equation}
C\left(\epsilon\right)=2\epsilon \lambda\label{eq:ce}
\end{equation}
and the inner and outer solutions are
\begin{align}
W&\sim\overbrace{2\mathrm{Im}\left\{ \frac{2}{\pi}\mathrm{cos}^{-1}\left[\frac{\mathrm{cos}\left(\pi \Theta_{\parallel}/2\right)}{\mathrm{cos}\left(\pi\delta/2\right)}\right]\right\}}^{W_1}\epsilon+\overbrace{\left(-Y^{2}\right)}^{W_2}\epsilon^{2}+\mathrm{e.s.t.} \label{eq:pivf}\\
\tilde{w}&\sim \overbrace{-\tilde{y}^{2}
+2\tilde{y}}^{\tilde{w}_0}+\overbrace{2\lambda}^{\tilde{w}_1}\epsilon+\mathrm{e.s.t.},\label{eq:vouter}
\end{align}
respectively. Note that we did not have to expand $W$ (or $\widehat{W}$) in powers of $\epsilon$ to arrive at the above inner solution.

\subsubsection{Composite Solution\label{sub:Hydrodynamic-g-Solution}}
The solutions for the inner and outer regions are in agreement in the
overlap region, where the outer one keeps its form. Therefore, the inner solution, Eq.\,(\ref{eq:pivf}), which is accurate to all algebraic orders in $\epsilon$, is uniformly valid throughout the domain, i.e., it's the composite solution as per, in outer variables
\begin{equation}
\tilde{w}_{\mathrm{comp}}\sim -\tilde{y}^2 +2\epsilon\mathrm{Im}\left\{\frac{2}{\pi}\mathrm{cos}^{-1}\left[\frac{\mathrm{cos}\left(\pi \left(\tilde{x}/\epsilon + \mathrm{i}\tilde{y}/\epsilon\right)/2\right)}{\mathrm{cos}\left(\pi\delta/2\right)}\right]\right\}+\mathrm{e.s.t}. \label{eq:inhy}
\end{equation}

\subsubsection{Slip Length}
The  slip length is the metric by which the flow enhancement provided by texturing a microchannel with a SH surface(s) is quantified.  It's related to
the one-dimensional (Navier slip) problem, which does not resolve the local velocity field in the inner region, and is governed by
\begin{equation}
\frac{\mathrm{d}^2 \tilde{w}_{\mathrm{1d}}}{\mathrm{d} \tilde{y}^2}= -2,
\end{equation}
where ${\tilde{w}_{1\mathrm{d}}}$ is the dimensionless velocity averaged over a the period of the ridges. It's subject to 
the symmetry condition at $\tilde{y} = 1$ and a Navier-slip boundary condition on the SH surface as per
\begin{equation}
\tilde{w}_{1\mathrm{d},\tilde{y}=0}=\tilde{b}\left.\frac{\mathrm{d}\tilde{w}_{\mathrm{1d}}}{\mathrm{d} \tilde{y}}\right|_{\tilde{y}=0}\label{eq:hsd},
\end{equation}
where 
\begin{equation}
\tilde{b}=\frac{b}{h},
\end{equation}
is the dimensionless slip length. It follows that
\begin{equation}
\tilde{w}_{\mathrm{1d}}=-\tilde{y}^2+2\tilde{y}+2\tilde{b},
\label{eq:b1d}
\end{equation}
and its mean value is
\begin{equation}
\bar{\tilde{w}}_{\mathrm{1d}}=2\tilde{b} +\frac{2}{3}.
\label{eq:w1dpar}
\end{equation}

Following \citet{Lauga-03}, we compute the slip length by equating the mean velocities from the composite and one-dimensional (Navier slip) velocity profiles.  As with the flow configurations resolved by \citet{Philip-72,Philip-72b}, the velocity field for our configuration is the sum of that for one-dimensional, Poiseuille flow in a microchannel with smooth walls and a velocity-increment
field. The latter is not subject to any forcing term and obeys the shear-free boundary condition at $\tilde{y}=1$. Therefore, there is no net momentum transferred across any plane normal to the composite interface in the velocity-increment problem and thus the mean velocity increment
over the width of the domain is a constant, i.e., $2\epsilon \lambda+\mathrm{e.s.t.}$ 
Thus, the outer solution is valid throughout the domain insofar as computing the slip length and it follows from  (\ref{eq:Outer_velo_parallel}), (\ref{eq:ce}) and~(\ref{eq:b1d}) that 
\begin{equation}
\tilde{b}\sim \epsilon \lambda+\mathrm{e.s.t}.
\label{eq:bpar}
\end{equation}

\subsubsection{Discussion}
We utlized a matched asymptotic expansion to develop the preceding hydrodynamic results, as done by
\citet{Hodes-17}, \citet{Kirk-18} and \citet{Kirk-20}, and they faciliate the below solution of 
the thermla problem. \citet{Teo-09} obtained it by taking the limit as $\epsilon \rightarrow 0$ of the dual-series equations satisfying the mixed boundary condition and their numerical solutions of them accommodate arbitrary values of $\epsilon$ and $\delta$. \citet{Kirk-17} too showed that the error in our result is exponentially small. More recently, \citet{Marshall-17} obtained an exact formulae for the slip length (expressed as contour integrals) for arbitrary 
$\epsilon$ and $\delta$ and, additionally, weakly-curved menisci, by using conformal maps, loxodromic function theory and reciprocity arguments. Additional effects have been considered by many investigators. For example, \citet{Kirk-18} relaxed the assumption of weakly-curved menisci in the large solid fraction limit for $\epsilon \ll 1$. Moreover, in numerical studies, \citet{Game-17} captured edge and subphase effects and \citet{Game-19} captured inertial effects due to slowly-varying curvature along a shear-free meniscus on account of the pressure difference across the meniscus decreasing in the streamwise direction.

We note that when the microchannel is textured on only one side, the same analysis done by \citet{Hodes-17} yields the composite velocity profile as
\begin{equation}
\tilde{w}_{\mathrm{comp}}\sim -\tilde{y}^2 +\frac{\epsilon}{1+\epsilon \lambda} \mathrm{Im}\left\{\frac{2}{\pi}\mathrm{cos}^{-1}\left[\frac{\mathrm{cos}\left(\pi \left(\tilde{x}/\epsilon + \mathrm{i}\tilde{y}/\epsilon\right)/2\right)}{\mathrm{cos}\left(\pi\delta/2\right)}\right]\right\}
+\mathrm{e.s.t}. \label{eq:inhys}
\end{equation}
 Also, an exact solution
based on a (cumbersome) conformal map is available for this case \citep{Philip-72,Philip-72b} and comparison with it shows that the composite velocity profile and corresponding slip length are accurate to $O\left(\mathrm{e}^{-\pi/\epsilon}\right)$ \citep{Hodes-17}, confirming that the error is exponentially small.

\subsection{Thermal Problem\label{sub:Thermal-Asymptotic-Analysis}}
\subsubsection{Formulation}
The dimensional form of the thermal energy equation is 
\begin{equation}
w\frac{\partial T}{\partial z}=\alpha\left(\frac{\partial^{2}T}{\partial x^{2}}+\frac{\partial^{2}T}{\partial y^{2}}\right),\label{eq:Therm_energy_dimensional_unsimp}
\end{equation}
where axial conduction is absent because the temperature field is fully developed and $\alpha=k/\left(\rho c_{\mathrm{p}}\right)$ is the thermal
diffusivity of the liquid, where $k$ is its thermal conductivity, $\rho$
is its density and $c_{\mathrm{p}}$ is its specific heat at constant pressure. The boundary conditions on Eq.~(\ref{eq:Therm_energy_dimensional_unsimp}) are 
\begin{align}
\frac{\partial{T}}{\partial{y}}&=0\,\,\,\mathrm{at}\,\,\,{y}=0\,\,\,\mathrm{for}\,\,\,0<{x}<a \\
\frac{\partial{T}}{\partial{y}}&=-\frac{q''_{\mathrm{sl}}}{k}\,\,\,\mathrm{at}\,\,\,{y}=0\,\,\,\mathrm{for}\,\,\,a<{x}< d \\
\frac{\partial{T}}{\partial{y}}&=0\,\,\,\mathrm{at}\,\,\,{y}=H\,\,\,\mathrm{for}\,\,\,0<{x}<d \\
\frac{\partial{T}}{\partial{x}}&=0\,\,\,\mathrm{at}\,\,\,{x}=0\,\,\,\mathrm{and}\,\,\,{x}=a\,\,\,\mathrm{for}\,\,\,0<{y}<H,
\end{align}
where $q''_{\mathrm{sl}}$ is the (constant) heat flux prescribed along the solid-liquid interface.
Moreover, again, because of the thermally-fully developed assumption, along with the constant heat flux boundary condition,
we have
\begin{equation}
\frac{\partial T}{\partial z}=\frac{q_{\mathrm{sl}}''(d-a)}{Q\rho c_{\mathrm{p}}},
\end{equation}
where $Q$ is the volumetric flow rate of liquid through a half period.

Henceforth, we proceed with the non-dimensional form of the problem, where temperature is non-dimensionalized (in a similar way as was done by \citet{Kirk-17}) by subtracting the (bulk) mean liquid temperature, $T_{\mathrm{m}}$, and dividing by a characteristic temperature scale  of $q_{\mathrm{sl}}''H/k$ such that
\begin{equation}
\tilde{T}=\frac{k\left(T-T_{\mathrm{m}}\right)}{q_{\mathrm{sl}}''H}\label{eq:ndtm},
\end{equation}
where
\begin{equation}
T_{\mathrm{m}}=\frac{\int_0^H \int_0^dwT\,\mathrm{d}x\,\mathrm{d}y}{\int_0^H \int_0^dw\,\mathrm{d}x\,\mathrm{d}y}.\label{eq:mixedm}
\end{equation}

The  thermal energy equation becomes
\begin{equation}
\frac{\epsilon \phi}{\tilde{Q}}\tilde{w}=\frac{\partial^{2}\tilde{T}}{\partial\tilde{x}^{2}}+\frac{\partial^{2}\tilde{T}}{\partial\tilde{y}^{2}},
\end{equation}
where $\phi=1-\delta$ is the solid fraction of the ridges and 
$\tilde{Q} = 2\mu Q/\left[\left(-\mathrm{d}p/\mathrm{d}z\right) H^4\right]$. 
The forcing term, $\epsilon \phi \tilde{w}/\tilde{Q}$, is known (to all algebraic orders) from the solution to the hydrodynamic problem.

Noting that
\begin{equation}
\tilde{Q}\sim \left(2\epsilon \lambda +2/3\right)\epsilon+\mathrm{e.s.t.},~~~\epsilon \rightarrow 0,
\end{equation}
the dimensionless thermal energy equation (written in terms of $X$ such that the domain boundaries are independent
of $\epsilon$) becomes 
\begin{equation}
\frac{\phi}{2/3+2\epsilon \lambda}\tilde{w}+\mathrm{e.s.t.}=\frac{1}{\epsilon^{2}}\frac{\partial^{2}\tilde{T}}{\partial X^{2}}+\frac{\partial^{2}\tilde{T}}{\partial\tilde{y}^{2}}\label{eq:Parallel outer thermal energy eq}
\end{equation}
subject to 
\begin{align}
\frac{\partial{\tilde{T}}}{\partial{\tilde{y}}}&=0\,\,\,\mathrm{at}\,\,\tilde{y}=0\,\,\,\mathrm{for}\,\,\,0<{X}<\delta \\
\frac{\partial{T}}{\partial{\tilde{y}}}&=-1\,\,\,\mathrm{at}\,\,\,\tilde{y}=0\,\,\,\mathrm{for}\,\,\,\delta<{X}< 1 \\
\frac{\partial\tilde{T}}{\partial\tilde{y}}&=0\,\,\,\mathrm{at}\,\,\,\tilde{y}=1\,\,\,\mathrm{for}\,\,\,0<{X}<1 \\
\frac{\partial\tilde{T}}{\partial{X}}&=0\,\,\,\mathrm{at}\,\,\,{X}=0\,\,\,\mathrm{and}\,\,\,{X}=1.
\end{align}

\subsubsection{Outer Region}
The forcing term in the above problem is $O\left(1\right)$ and thus by the same reasoning used to justify that $\tilde{w} =\tilde{w}\left(\tilde{y}\right) = O\left(1\right)$, we also have $\tilde{T}=\tilde{T}\left(\tilde{y}\right) = O\left(1\right)$ to leading order. 
We thus expand the temperature field as
\begin{equation}
\tilde{T}=\sum_{n=0}^{\infty}\epsilon^{n}\tilde{T}_{n}+\mathrm{e.s.t.},~~~\epsilon \rightarrow 0,\label{eq:pote}
\end{equation}
where $\tilde{T}_{n}=O(1)$ for all $n\geq0$. 

Substituting Eq.\,(\ref{eq:pote}) and the solution for the 
outer-velocity profile as per  (\ref{eq:vouter}) 
into Eq.\,(\ref{eq:Parallel outer thermal energy eq}), we find that, at the
various orders of $\epsilon$,

\begin{align}
O\left(\epsilon^{-2}\right):~~~&\frac{\partial^{2} \tilde{T}_{0}}{\partial X^2}=0 \\
O\left(\epsilon^{-1}\right):~~~&\frac{\partial^{2} \tilde{T}_{1}}{\partial X^2}=0 \\ 
O\left(\epsilon^{0}\right):~~~&\frac{\partial^{2} \tilde{T}_{2}}{\partial X^2}+\frac{\partial^{2} \tilde{T}_{0}}{\partial \tilde{y}^{2}}=\frac{3\phi}{2}\left(-\tilde{y}^{2}+2\tilde{y}\right) \\
O\left(\epsilon^{n}\right):~~~&\frac{\partial^{2} \tilde{T}_{n+2}}{\partial X^2}+\frac{\partial^{2} \tilde{T}_{n}}{\partial \tilde{y}^{2}}=\frac{3\phi}{2}\left[
\left(-3\lambda\right)^n\left(-\tilde{y}^2+2\tilde{y}\right)+\left(-3\lambda\right)^{n-1}2\lambda\right],~n\geq 1.
\end{align}
Following the same approach as in the outer hydrodynamic problem shows that  $\tilde{T}_{n}=\tilde{T}{}_{n}(\tilde{y})$ for all $n$.
The solution to the outer problem, which satisfies all 3 symmetry conditions, follows as
\begin{equation}
\tilde{T}=\frac{\phi}{2/3+2\epsilon \lambda}\left[-\frac{\left(\tilde{y}-1\right)^{4}}{12}+\left(\frac{1}{2}+\epsilon \lambda\right)\left(\tilde{y}-1\right)^{2}+D\left(\epsilon\right)\right]
+\mathrm{e.s.t.},~\epsilon \rightarrow 0,
\label{eq:poutT}
\end{equation}
where
\begin{equation}
D(\epsilon)=\sum_{n=0}^{\infty}\epsilon^{n}D_{n}
\end{equation}

\subsubsection{Inner Region}
Denoting the inner temperature profile as $\tilde{T}=\theta\left(X,Y\right)$, the thermal energy equation becomes
\begin{equation}
\frac{\partial^{2}\theta}{\partial X^{2}}+\frac{\partial^{2}\theta}{\partial Y^{2}}=\frac{\phi}{2/3+2\epsilon \lambda}\left(-\epsilon^{4}Y^2+2\epsilon^{3}\mathrm{Im}\left\{ \frac{2}{\pi}\mathrm{cos}^{-1}\left[\frac{\mathrm{cos}\left(\pi Z/2\right)}{\mathrm{cos}\left(\pi\delta/2\right)}\right]\right\} \right)+\mathrm{e.s.t.}\label{eq:pies}
\end{equation}
The forcing is $O\left(\epsilon^{3}\right)$; therefore, neglecting terms of this order, it reduces to Laplace's equation as per
\begin{equation}
\frac{\partial^{2}\theta}{\partial X^{2}}+\frac{\partial^{2}\theta}{\partial Y^{2}}=
O\left(\epsilon^3\right)
\label{eq:oi}
\end{equation}
subject to
\begin{align}
\frac{\partial\theta}{\partial Y}&=0\,\,\mathrm{at}\,\, Y=0\,\,\mathrm{for}\,\,0<X<\delta \\
\frac{\partial\theta}{\partial Y}&=-\epsilon\,\,\mathrm{at}\,\, Y=0\,\,\mathrm{for}\,\,\delta<X<1 \\
\theta&\sim-\epsilon\phi Y+\phi\left[\frac{1}{2}+\frac{1/12+D(\epsilon)}{2/3+2\epsilon \lambda}\right]+O\left(\epsilon^{3}\right)\,\,\mathrm{as}\,\, Y\rightarrow\infty \label{eq:tpm}\\
\frac{\partial\theta}{\partial X}&=0\,\,\mathrm{at}\,\, X=0\,\,\mathrm{and}\,\, X=1~\mathrm{for}~Y>0,
\end{align}
where  (\ref{eq:tpm}), which follows from  (\ref{eq:poutT}) with $\tilde{y} = \epsilon Y$, is the matching condition in the overlap region.
As per the solution by \citet{Mikic-57} to this problem in the context of (thermal) spreading resistance,
\begin{equation}
\theta = -\epsilon \phi Y  - \frac{2\epsilon}{\pi^2}\sum_{n=1}^{\infty}\frac{\sin\left(n\pi\delta\right)\cos\left(n\pi X\right)\mathrm{e}^{-n \pi Y}}{n^2}+\phi\left[\frac{1}{2}+\frac{1/12+D(\epsilon)}{2/3+2\epsilon \lambda}\right]+O\left(\epsilon^{3}\right).
\label{eq:mikic}
\end{equation}

We solve for $D\left(\epsilon\right)$ by enforcing that the dimensionless mean liquid temperature, as defined by  (\ref{eq:ndtm}) and~(\ref{eq:mixedm}), is zero, i.e., using outer variables,
\begin{equation}
\int_{0}^{1}\int_0^{\epsilon}\tilde{w}\tilde{T}\mathrm{d}\tilde{x}\mathrm{d}\tilde{y}=0.
\end{equation}
Next, we express this result as
\begin{equation}
\int_0^{\eta}\int_0^{\epsilon}\tilde{w}_{\mathrm{in}}\tilde{T}_{\mathrm{in}}\mathrm{d}\tilde{x}\mathrm{d}\tilde{y}+\int_0^1\int_0^{\epsilon}\tilde{w}_{\mathrm{out}}\tilde{T}_{\mathrm{out}}\mathrm{d}\tilde{x}\mathrm{d}\tilde{y}-\int_0^{\eta}\int_0^{\epsilon}\tilde{w}_{\mathrm{out}}\tilde{T}_{\mathrm{out}}\mathrm{d}\tilde{x}\mathrm{d}\tilde{y}=0,
\label{eq:tb3}
\end{equation}
where $\eta$ is a value of $\tilde{y}$ in the overlap region and nondimensional velocity and temperature have been subscripted for clarity. Denoting the (uniformly valid) composite hydrodynamic solution from  (\ref{eq:inhy}) in outer variables as 
$\tilde{w}_{\mathrm{in}}$ and the inner thermal solution from  (\ref{eq:mikic}) in them by $\tilde{T}_{\mathrm{in}}$, we write the first term in this equation as
\begin{align}
\int_0^{\eta}\int_0^{\epsilon}\tilde{w}_{\mathrm{in}}\tilde{T}_{\mathrm{in}}\mathrm{d}\tilde{x}\mathrm{d}\tilde{y} =&\phi\left[\frac{1}{2}+\frac{1/12+D(\epsilon)}{2/3+2\epsilon \lambda}\right]\int_0^{\eta}\int_0^{\epsilon}\tilde{w}_{\mathrm{in}}\mathrm{d}\tilde{x}\mathrm{d}\tilde{y} - \phi \int_0^{\eta}\int_0^{\epsilon}\tilde{y}\tilde{w}_{\mathrm{in}} \mathrm{d} \tilde{y} 
\nonumber \\
&+  O\left(\epsilon^3\eta\right) \label{eq:tb2}.
\end{align}
Next, we observe that the solution to the $\widehat{W}$ problem is that for a one-dimensional (Couette flow) problem, i.e., $Y + C\left(\epsilon\right)/\left(2\epsilon\right)$, plus that to a perturbation problem with a mean velocity of zero along any plane spanning the width of the domain. It then follows from  (\ref{eq:Outer_velo_parallel}) and (\ref{eq:Inner_velo_incomp}) that we can replace $\int_0^{\epsilon}\tilde{w}_{\mathrm{in}}\mathrm{d}\tilde{x}$ with $\int_0^{\epsilon}\tilde{w}_{\mathrm{out}}\mathrm{d}\tilde{x}$ in  (\ref{eq:tb2}). Then,  none of the integrands in  (\ref{eq:tb3}) depend on $\tilde{x}$; consequently, it becomes
\begin{align}
\phi\left[\frac{1}{2}+\frac{1/12+D(\epsilon)}{2/3+2\epsilon \lambda}\right]\int_0^{\eta}
\tilde{w}_{\mathrm{out}}\mathrm{d}\tilde{y}-  \phi \int_0^{\eta}\tilde{y}\tilde{w}_{\mathrm{out}} \mathrm{d} \tilde{y} &\nonumber \\
+\int_0^{1}\tilde{w}_{\mathrm{out}}\tilde{T}_{\mathrm{out}}\mathrm{d}\tilde{y}-\int_0^{\eta}\tilde{w}_{\mathrm{out}}\tilde{T}_{\mathrm{out}}\mathrm{d}\tilde{y}+ O\left(\epsilon^2\eta\right) &= 0 \label{eq:tb4}.
\end{align}
Moreover, the first, second and fourth terms in  (\ref{eq:tb4}) sum to error terms that are negligible  compared to 
$O\left(\epsilon^2\eta\right)$ such that it simplifies to
\begin{equation}
\int_0^1\tilde{w}_{\mathrm{out}}\tilde{T}_{\mathrm{out}}\mathrm{d}\tilde{y}=O\left(\epsilon^2 \eta\right).
\end{equation}
The error estimate applies for any choice of $\eta = \epsilon^{\alpha}$ in the overlap region; therefore, it's smaller than $\epsilon^{2+\alpha}$ for $0<\alpha<1$, implying that $\int_0^1\tilde{w}_{\mathrm{out}}\tilde{T}_{\mathrm{out}}\mathrm{d}\tilde{y}= O\left(\epsilon^3\right)$, i.e.,
\begin{equation}
\int_0^1 \left(-\tilde{y}^2+2\tilde{y}+2\lambda\epsilon\right)
\frac{\phi}{2/3+2\epsilon \lambda}\left[-\frac{\left(\tilde{y}-1\right)^{4}}{12}+\left(\frac{1}{2}+\epsilon \lambda\right)\left(\tilde{y}-1\right)^{2}+D\left(\epsilon\right)\right]\mathrm{d}\tilde{y} = O\left(\epsilon^3\right).
\label{eq:bulkint}
\end{equation}
Evaluating the integral, we find that
\begin{equation}
D\left(\epsilon\right) =\breve{D}\left(\epsilon\right)+O\left(\epsilon^3\right),
\end{equation}
where
\begin{equation}
\breve{D}\left(\epsilon\right) = -\frac{1}{140}\frac{13+91\epsilon \lambda + 140 \left(\epsilon \lambda\right)^2}{1+3\epsilon \lambda},
\label{eq:def}
\end{equation}
which resolves the inner temperature profile as per  (\ref{eq:inth}). 

We do not expand the right side of  (\ref{eq:def}) for small $\epsilon \lambda$ because, 
if we take a secondary limit where the solid fraction of the ridges approaches zero (a relevant limit in practice to maximize lubrication), $\lambda \rightarrow \infty$; therefore, expansion for small $\epsilon \lambda$ will clearly break down. However, the penultimate term on the right side of  (\ref{eq:mikic}) is well 
behaved in this limit and it follows from  (\ref{eq:def}) that 
\begin{equation}
\frac{1}{2}+\frac{1/12+D(\epsilon)}{2/3+2\epsilon \lambda}=\hat{D}\left(\epsilon \lambda\right) +O\left(\epsilon^3\right).
\label{eq:dhatdef}
\end{equation}
where
\begin{equation}
\hat{D}\left(\epsilon \lambda\right) = \frac{17+84\epsilon \lambda + 105 \left(\epsilon \lambda\right)^2}
{35\left(1+3\epsilon \lambda\right)^2},
\label{eq:dhat}
\end{equation}
which varies between 1/3 and 17/35 for arbitrary values of $\epsilon \lambda$.
It then follows from  (\ref{eq:mikic}) that the temperature along the composite interface is
\begin{equation}
\theta \left(X,0\right) = \phi \hat{D}\left(\epsilon \lambda\right)  - \frac{2\epsilon}{\pi^2}\sum_{n=1}^{\infty}\frac{\sin\left(n\pi\delta\right)\cos\left(n\pi X\right)}{n^2}+O\left(\epsilon^{3}\right),
\label{eq:thetacp}
\end{equation}

\subsubsection{Composite Solution}
The inner solution is not the composite one as in the hydrodynamic problem because  (\ref{eq:tpm}) is consistent with  (\ref{eq:poutT}) only up to $O\left(\epsilon^2\right)$. Rather,
\begin{equation}
\tilde{T}_{\mathrm{comp}} = \tilde{T}_{\mathrm{out}} 
+ \tilde{T}_{\mathrm{in}} - \tilde{T}_{\mathrm{overlap}};
\label{eq:tcomp}
\end{equation}
consequently,
\begin{align}
\tilde{T}_{\mathrm{comp}} &\sim \frac{\phi}{\frac{2}{3}+2\epsilon \lambda}\left[-\frac{\left(\tilde{y}-1\right)^{4}}{12}+\left(\frac{1}{2}+\epsilon \lambda\right)\left(\tilde{y}-1\right)^{2}+\breve{D}\left(\epsilon\right)\right]
 \nonumber
 \\
 &- \frac{2\epsilon}{\pi^2}\sum_{n=1}^{\infty}\frac{\sin\left(n\pi\delta\right)\cos\left(n\pi \frac{\tilde{x}}{\epsilon}\right)\mathrm{e}^{-n \pi \frac{\tilde{y}}{\epsilon}}}{n^2}+O\left(\epsilon^3\right).
 \label{eq:tcomp}
\end{align}
\subsubsection{Nusselt Numbers}
We define the local Nusselt number along the solid-liquid interface as 
\begin{equation}
\mathrm{Nu}(X) = \frac{4h(X)H}{k},
\label{eq:nu}
\end{equation}
where $h(X)$ is the (local) heat transfer coefficient, i.e., 
$q''_{\mathrm{sl}}/\left(T_{\mathrm{sl}}-T_{\mathrm{m}}\right)$, which is finite along the solid-liquid interface and zero along the meniscus. Then,
\begin{equation}
\mathrm{Nu}(X) = 
\begin{cases}
0~~~\mathrm{for}~~~0\leq X < \delta\\
\frac{4}{\theta\left(X,0\right)}~~~\mathrm{for}~~~\delta < X \leq 1.
\end{cases}
\label{eq:nudef}
\end{equation}
The minimum local Nusselt number ($\mathrm{Nu_{min}}$) occurs at the center of a ridge, i.e.,  $X=1$, as
per
\begin{equation}
\mathrm{Nu_{min}} = \frac{4}{\theta\left(1,0\right)}.
\label{eq:nmdef}
\end{equation}
This is an important engineering parameter because the magnitude of the maximum temperature difference between the ridge and bulk liquid follows from it.
From  (\ref{eq:thetacp}),
the minimum local Nusselt number ($\mathrm{Nu_{min}}$)  may be expressed as
\begin{equation}
\mathrm{Nu_{min}} =
\frac{4}{ \phi  \left[ \hat{D}\left(\epsilon \lambda\right)+\frac{2\epsilon}{\phi}
\sum_{n=1}^{\infty}\frac{\sin\left(n\pi\phi\right)}{\left(n\pi\right)^2}\right]}+O\left(\epsilon^{3}\right).
\label{eq:numinKp}
\end{equation}


More generally, the local Nusselt number anywhere along the ridge is
\begin{equation}
\mathrm{Nu}(X) =
\frac{4}{ \phi \left( \hat{D}\left(\epsilon \lambda\right)  - \frac{2\epsilon}{\phi}
\sum_{n=1}^{\infty}\frac{\sin\left(n\pi\delta\right)\cos\left(n\pi X\right)}{n^2\pi^2}\right) }+O\left(\epsilon^{3}\right), \quad \delta < X \leq 1.
\label{eq:nuloc}
\end{equation}
This was derived in the limit of small $\epsilon$, but the subsequent behaviour in the small solid fraction limit, $\phi \to 0$, is also of interest. This limit is physically relevant, but care should be taken when making further approximations in $\epsilon$ that the correct asymptotic behaviour in the secondary limit $\phi \to 0$ is preserved. As we will show, this will considerably increase the accuracy of any resulting approximation. First, note that the sum in  (\ref{eq:nuloc}) which we denote by $S(X)$, can be written
\begin{align}
	S(X) = \frac{2}{\phi}
\sum_{n=1}^{\infty}\frac{\sin\left(n\pi\delta\right)\cos\left(n\pi X\right)}{n^2\pi^2} &= -2 \sum_{n=1}^\infty \frac{\cos(n\pi(1-X))}{n\pi} + o(1), \quad \phi \to 0 \nonumber \\
	&= \frac{2}{\pi}\ln\left|2\sin\left(\frac{\pi(1-X)}{2}\right)\right| + o(1),
\end{align}
where the last equality follows from the complex representation of cosine (i.e. $\cos\theta = (\mathrm{e}^{\mathrm{i}\theta} + \mathrm{e}^{-\mathrm{i}\theta})/2$), and the Taylor series of $-\log (1-k)$ around $k=0$. As $X$ is restricted to the ridge, substituting $X = 1-\phi t$ where $0\leq t < 1$, and expanding for $\phi \to 0$, we find
\begin{align}
	S(X) &= \frac{2}{\pi}\ln\phi + \frac{2}{\pi}\ln(\pi t) + o(1),\quad \phi \to 0, \quad 0\leq t<1.
\end{align}
Therefore, $S$ has a (constant in $X$ or $t$) logarithmic singularity as $\phi \to 0$. Additionally, $\lambda \sim -(2/\pi)\ln \phi$ has a similar logarithmic singularity. Therefore, substituting into (\ref{eq:nuloc}) and using that $\hat{D} \to 1/3$ as $\phi \to 0$,
\begin{align}
\mathrm{Nu} &\sim -\frac{2\pi}{ \epsilon \phi\ln\phi  },~~~\phi \to 0,
\end{align}
anywhere on the ridge, $\delta< X \leq 1$. In particular, $\mathrm{Nu_{min}} \to \infty$ which is expected physically.

If we had expanded (\ref{eq:nuloc}) in a regular power series in $\epsilon$, without the knowledge of $S$ and $\lambda$ for small $\phi$, we would have
\begin{align}
	\mathrm{Nu}(X) &= \frac{140}{17\phi}\left(1-\frac{18}{17}\epsilon \lambda - \frac{35}{17}\epsilon S(X) + O(\epsilon^2)\right), \quad \delta<X\leq 1 \label{eq:nuloc_bad}
\end{align}
and although valid for fixed $\phi$, as $\phi \to 0$ the logarithmic singularities mean $\epsilon\lambda$ and $\epsilon S \to \infty$, and the expansion breaks down. Thus, the expansion (\ref{eq:nuloc_bad}) should be avoided. The unexpanded expression (\ref{eq:nuloc}) is already explicit, but to facilitate averaging, an expansion which does not break down as in the small $\phi$ limit is desirable. To achieve this, we can factor out this logarithmic behaviour. First, add and subtract $\epsilon \lambda$ from the denominator of (\ref{eq:nuloc}), giving  
\begin{align}
	\mathrm{Nu}(X) &=
\frac{4}{ \phi \left[\hat{D}\left(\epsilon \lambda\right) +\epsilon \lambda -\epsilon (\lambda + S(X)) \right] }+O\left(\epsilon^{3}\right), \quad \delta<X\leq 1.
\end{align}
Then, factor out $\hat{D}\left(\epsilon \lambda\right) +\epsilon \lambda$ and define $\mathcal{E}(\epsilon)= \epsilon / (\hat{D} + \epsilon \lambda)$,
\begin{align}
	\mathrm{Nu}(X) &=
\frac{4}{ \phi \left[\hat{D}\left(\epsilon \lambda\right) +\epsilon \lambda\right]\left[1  -\mathcal{E}(\epsilon) (\lambda + S(X)) \right] }+O\left(\epsilon^{3}\right), \quad \delta<X\leq 1.
\end{align}
Now, it can be shown that $0<\mathcal{E}(\epsilon) < 35 \epsilon /17$ for any $\phi$, and hence it remains $O(\epsilon)$ as $\epsilon \to 0$, even if $\phi \to 0$ simultaneously. Additionally, $\lambda + S(X)$ remains bounded in the limit $\phi \to 0$. Taylor expanding the denominator results in a well-ordered series,
\begin{align}
\mathrm{Nu}(X) =& \frac{4}{\phi \left[\hat{D}\left(\epsilon \lambda\right)+\epsilon \lambda\right]}
\left\{1+ \mathcal{E}(\epsilon)\left[ \lambda+S(X)\right] 
+ \mathcal{E}(\epsilon)^2\left[\lambda+S(X)\right]^2\right\} + O\left(\epsilon^3\right) \nonumber \\
& \mathrm{for}~\delta < X \leq 1, \label{eq:nulocal_expanded}
\end{align}
where the $O(\epsilon^3)$ error term accounts for the $O(\mathcal{E}^3)$ terms since $\mathcal{E} = O(\epsilon)$.
We note that by evaluating Nu in the limit when the solid fraction is 1, i.e., for a smooth channel such that $\lambda \rightarrow 0$, we recover the well-known result that Nu = 140/17. (The $O\left(\epsilon^3\right)$ error term can be shown to vanish in this limit.) 

The average Nusselt number for the composite interface is defined in the manner of \citet{Maynes-14} and \citet{Kirk-17} as
\begin{equation}
\overline{\mathrm{Nu}} = \frac{4\overline{h}H}{k},
\label{eq:barNu1}
\end{equation}
where $\overline{h} = \int_a^d h\mathrm{d}x/d$. Thus,
\begin{equation}
\overline{\mathrm{Nu}}=\int_{\delta}^{1}\mathrm{Nu}(X)\,\mathrm{d}X.
\label{eq:barNu}
\end{equation}  
Performing the integration on (\ref{eq:nulocal_expanded}), and substituting the series $S(X)$, we find that
\begin{align}
\overline{\mathrm{Nu}} &\sim \frac{4}{\hat{D}\left(\epsilon \lambda\right)+\epsilon \lambda}
\left\{
1+\left[
\lambda-\sum_{n=1}^{\infty}\frac{2\sin^2\left(n\pi\delta\right)}{\phi^2\left(n\pi\right)^3}
\right]\mathcal{E}(\epsilon)
\right. \nonumber \\
& \left.+
\left[\lambda^2-2\lambda\sum_{n=1}^{\infty}\frac{2\sin^2\left(n\pi\delta\right)}{\phi^2\left(n\pi\right)^3}+\sum_{m=1}^{\infty}\sum_{n=1}^{\infty}L_{m,n}\right]\mathcal{E}(\epsilon)^2\right\}
+O\left(\epsilon^3\right),
\label{eq:nupf}
\end{align}
where
\begin{equation}
L_{mn} = \frac{4}{\phi^3}\frac{\sin\left(n\pi\delta\right)
\sin\left(m\pi\delta\right)}{\left(n\pi\right)^2\left(m\pi\right)^2}
\begin{cases}
\frac{1}{2}\left[1-\delta - \frac{\sin\left(2n\pi\delta\right)}{2n\pi}\right]~\mathrm{for}~m=n\\
-\frac{\sin\left[\left(m+n\right)\pi\delta\right]}{2\pi\left(m+n\right)} - \frac{\sin\left[\left(m-n\right)\pi\delta\right]}{2\pi\left(m-n\right)}
~\mathrm{for}~m \neq n.
\end{cases}
\end{equation}
Alternatively, a simpler expression valid for any solid fraction and no sums to evaluate, but with an $O\left(\epsilon\right)$
error term, follows from the leading term of  (\ref{eq:nupf}),
\begin{equation}
\overline{\mathrm{Nu}} \sim \frac{4}{\hat{D}\left(\epsilon \lambda\right)+\epsilon \lambda} + O\left(\epsilon\right).
\label{eq:nupfs}
\end{equation}

An average Nusselt number may also be defined in the manner of \citet{Enright-14}
as
\begin{equation}
\overline{\mathrm{Nu}}' =\frac{4\bar{h}'H}{k}
\label{eq:nubar2}
\end{equation}
 by defining the average heat transfer coefficient as $\bar{h}'= \phi q''_{\mathrm{sl}}/(\overline{T_{\mathrm{sl}}-T_{\mathrm{m}}})$ such that it's based upon the mean heat flux over the composite 
interface ($\phi q''_{\mathrm{sl}}$) and the mean driving force for heat transfer over the solid-liquid one ($\overline{T_{\mathrm{sl}}-T_{\mathrm{m}}}$). Then, it's given by
\begin{equation}
\overline{\mathrm{Nu}}' = \frac{4\phi^2}{\int_{\delta}^1 \theta \left(X,0\right) \mathrm{d}X}.
\end{equation}
Performing the integration we find that
\begin{align}
\overline{\mathrm{Nu}}'= &140\left(1+3\epsilon \lambda\right)^2/\left\{
17+ \left[84\lambda + \frac{70}{\phi^2\pi^3}\sum_{n=1}^{\infty}\frac{\sin^2\left(n\pi\delta\right)}{n^3}\right]\epsilon \right. \nonumber \\
&\left.
 +\left[
105\lambda^2 +\frac{420\lambda}{\phi^2\pi^3}\sum_{n=1}^{\infty}\frac{\sin^2\left(n\pi\delta\right)}{n^3}
\right]\epsilon^2+\left[ \frac{630\lambda^2}{\phi^2\pi^3}\sum_{n=1}^{\infty}\frac{\sin^2\left(n\pi\delta\right)}{n^3}\right]\epsilon^3\right\} +O\left(\epsilon^3\right).\label{eq:nupp}
\end{align}
There is no need to expand this result further in $\epsilon$,
and it also vanishes as expected when $\lambda \rightarrow \infty$ (i.e., $\phi \rightarrow 0$). Also, the last term in the denominator is retained as, depending on the relative values of $\lambda$ and $\epsilon$, it may be larger than 
$O\left(\epsilon^3\right)$. 
Finally, we note that, regardless of which of the two definitions for the mean Nusselt number is used, the local one is given by  (\ref{eq:nu}).

\subsubsection{Validation}
\begin{figure}
\begin{centering}
\includegraphics[width=9cm]{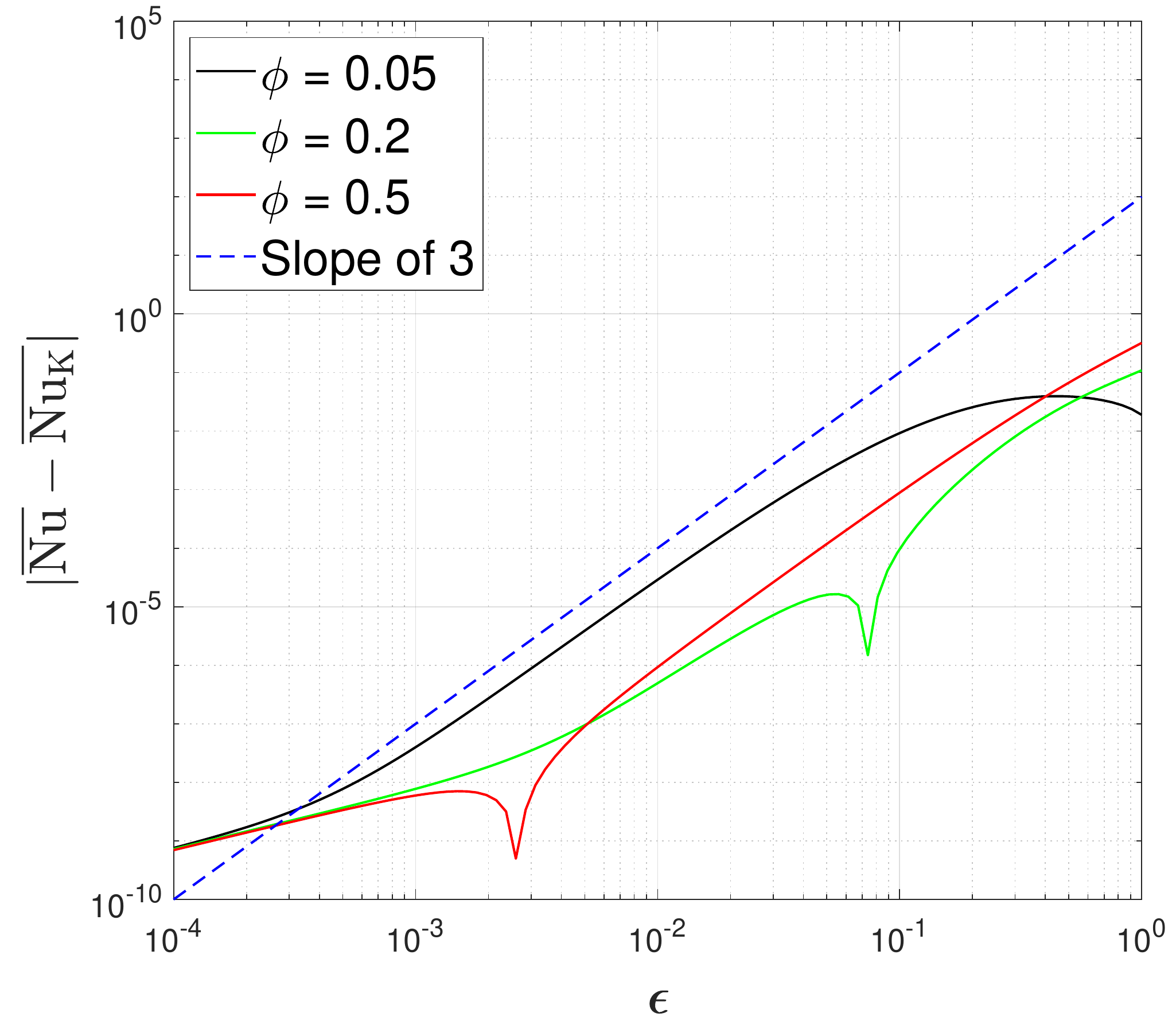}
\par\end{centering}
\caption{Log-log plot of error in $\overline{\mathrm{Nu}}$ as per  (\ref{eq:nupf}) relative to exact result 
from \citet{Kirk-17} versus $\epsilon$ for $\phi$ = 1/20, 1/5 and 1/2.}
\label{fig:error}
\end{figure}

\citet{Kirk-17} solved the problem exactly using eigenfunction expansions, whereby  the dual-series equations resulting from the mixed  boundary condition on the hydrodynamic problem are numerically resolved. We validate our result for $\overline{\mathrm{Nu}}$ as per  (\ref{eq:nupf}) via the log-log plot of its error relative to the exact result
(i.e., $\left|\overline{\mathrm{Nu}} - \overline{\mathrm{Nu}}_{\mathrm{K}}\right|$, where the subscript ``K'' denotes the
exact result by \citet{Kirk-17}) versus $\epsilon$ when $\phi$ = 1/20, 1/5 and 1/2 in
Fig.~\ref{fig:error}. The 
parameters used to generate this plot were as follows. For the present study, all (4) sums required to compute 
$\overline{\mathrm{Nu}}$ from  \eqref{eq:nupf} were truncated at 50,000. To compute  
$\overline{\mathrm{Nu}}_{\mathrm{K}}$ for the 
\citet{Kirk-17} study, the following parameters were used. First, to resolve the velocity field, we truncated the sums in the dual-series equations resulting from the mixed boundary condition at 800.\footnote{See  (4.10) and (4.11) in \citet{Kirk-17} and the discussion below them.} Secondly, in the expression corresponding to the surface temperature along the composite 
interface as per  (5.15) in \citet{Kirk-17}, we, by necessity, truncated the sums 
dependent upon the perturbation to the velocity field of a smooth channel at 800 and the 
remaining sum at 1.5 $\times$ 10$^6$. (The latter sum equals 0 in the limit as a smooth
channel is approached, i.e., $\phi \rightarrow 1$.) We used 800 points along the ridge to evaluate its surface temperature as required to compute $\overline{\mathrm{Nu}}_{\mathrm{K}}$. Observe that, as $\epsilon$ is decreased, all 3 curves in Fig. 2 approach a slope of 3, indicating that $|\bar{\mathrm{Nu}}-
\bar{\mathrm{Nu}}_{\mathrm{K}}| = O(\epsilon^3)$, as expected. When $\epsilon$ is sufficiently small ($\epsilon \ll 10^{-3}$), the slopes transition from 3 to 1. This is due to the unavoidable error introduced by truncation of the sums appearing at lower orders in $\epsilon$ in  \eqref{eq:nupf} and in the series solution of Kirk et al. Hence, the slope of 1 is due to truncation error in sums at $O(\epsilon)$ in both the asymptotic and exact solutions, and this transition to a slope of 1 can be moved to arbitrarily small values of $\epsilon$ by increasing the number of terms in all sums.

\subsubsection{Prior Results}
The prior studies by \citet{Maynes-14} and \citet{Kirk-17} provide the local Nusselt number 
as per  (\ref{eq:nu}) and, by implication, the minimum one as per  (\ref{eq:nmdef}), as well as the
mean Nusselt Number $\overline{\mathrm{Nu}}$ as per  (\ref{eq:barNu1}). The mean Nusselt number
$\overline{\mathrm{Nu}}'$ as per  (\ref{eq:nubar2}) also follows from the temperature profiles developed in these studies. 
Moreover, the studies by \citet{Enright-06} and \citet{Enright-14} provide $\overline{\mathrm{Nu}}'$ and machinery that may be adopted to find the local Nusselt numbers. We proceed to discuss and contrast all of them and our own results.

\citet{Maynes-14} used a Navier-slip velocity profile in the thermal energy equation. An analytical solution 
for the temperature profile was found by representing the discontinuous (Neumann) boundary condition along the composite interface as a Fourier series and using separation of variables. Their temperature profile (in our notation) along the composite interface is 
\begin{equation}
\theta_\mathrm{M}\left(X,0\right) = \frac{17}{35}\phi -\frac{2\epsilon}{\pi^2}\sum_{n=1}^{\infty}\frac{\sin\left(n\pi\delta\right)\coth\left(n\pi/\epsilon\right)\cos\left(n\pi X\right)}{n^2} -\frac{18}{35}\frac{\phi \epsilon  \lambda}{1+3\epsilon \lambda} +\frac{6}{35}\frac{\phi\epsilon^2 \lambda^2}{\left(1+3\epsilon \lambda\right)^2},
\label{eq:tslM}
\end{equation}
where the subscript ``M'' denotes a result by \citet{Maynes-14}. 
Closed-form expressions for $\mathrm{Nu}_{\mathrm{M}}$ and $\mathrm{Nu}_{\mathrm{min,M}}$ follow and 
their behaviour is discussed by \citet{Maynes-14}. A closed form expression for 
$\overline{\mathrm{Nu}}'_{\mathrm{M}}$ also follows from integrating  (\ref{eq:tslM}). Numerical integration was used by \citet{Maynes-14} to compute 
$\overline{\mathrm{Nu}}_{\mathrm{M}}$. Although \citet{Maynes-14} did not quantify the error in their Nusselt number expressions due to using a Navier-slip velocity profile, we have shown that it is $O\left(\epsilon^3\right)$. 
Relatedly, upon noting that $\coth\left(n\pi/\epsilon\right) \sim 1 + \mathrm{e.s.t.}$ as $\epsilon \rightarrow 0$,  an expansion of  (\ref{eq:tslM}) is consistent with  (\ref{eq:thetacp}) and all of the Nusselt numbers become the same as in our analysis. We note, however, that, since \citet{Maynes-14} did not do this, their results differ slightly from ours in
the plots below. 

Insofar as $\overline{\mathrm{Nu}}'$, \citet{Enright-06} were the first to compute it and, as here, solid-liquid interfaces were isoflux. They too utilized a Navier-slip velocity profile as per  (\ref{eq:w1dpar}), or the corresponding expression when one side of the microchannel is textured, but did not capture the discontinuous thermal boundary condition along the composite interface. Consequently, the Nusselt number was solely dependent upon the slip length non-dimensionalized by the channel height for the texture of interest (parallel or transverse ridges, pillars, etc.). \citet{Enright-14} refined this approach by further imposing an apparent temperature jump along the composite (c) interfaces on one or both sides of a microchannel as per 
\begin{equation}
\bar{T}_{\mathrm{sl}} - \bar{T}_{\mathrm{c}} = -b_{\mathrm{t}} \left.\frac{\partial \bar{T}}{\partial n}\right|_{\mathrm{c}},
\label{eq:btd}
\end{equation}
where $b_{\mathrm{t}}$ is the apparent thermal slip length (also referred to as the temperature jump length) and $n$ the direction normal 
to a composite interface and pointing into the liquid. Using the approach of \citet{Nield-04,Nield-08} to accommodate asymmetrical boundary conditions, closed-form expressions for the Nusselt number were developed as a function of arbitrary values of $b$ and $b_{\mathrm{t}}$ imposed on each side of the microchannel and, additionally, the ratio of heat fluxes averaged over the composite interfaces ($\phi q''_{\mathrm{sl}}$) bounding the domain (liquid). In the case of a symmetric channel, as in this study, the results by \citet{Enright-14} reduce to those developed by \citet{Inman-64} 50 years earlier in the context of molecular slip effects in gas flows, where the surface boundary conditions are mathematically equivalent. \citet{Kane-17} extended the \cite{Enright-14} analysis to a combined Poiseuille and Couette flow.

The (one dimensional) temperature field provided by the foregoing studies in the case of parallel ridges is that averaged over the width of the domain.  Insofar as the definition of dimensionless temperature adopted here as per  (\ref{eq:ndtm}), only the composite interface temperature is relevant; consequently, the thermal slip length is irrelevant. The resulting temperature profile, which we denote by $\tilde{T}_{\mathrm{E}}$, is
\begin{equation}
\tilde{T}_{\mathrm{E}} = \frac{\phi}{2/3+2\epsilon \lambda}\left(
-\frac{\tilde{y}^4}{12} + \frac{\tilde{y}^3}{3} + \epsilon \lambda \tilde{y}^2 -2\epsilon \lambda \tilde{y} - \frac{2}{3} \tilde{y} 
+\frac{2}{105}\frac{17+84\epsilon \lambda + 105 \epsilon^2 \lambda^2}{1+3\epsilon \lambda}
\right).
\end{equation}
This expression is identical to that given by  (\ref{eq:poutT}) with $D\left(\epsilon\right)$ given by 
 (\ref{eq:def}).
Terms of $O\left(\epsilon^3\right)$ were neglected in our derivation and thus, albeit not pointed out by \citet{Enright-14}, it is only accurate to $O\left(\epsilon^3\right)$.

In the foregoing studies (except in \citet{Enright-06}), the difference between the mean temperatures of the solid-liquid and composite interfaces follows by imposing an apparent thermal slip boundary condition as per  (\ref{eq:btd}). \citet{Enright-14} utilized the spreading resistance solution by \citet{Mikic-57} to evaluate $b_{\mathrm{t}}$ such that
\begin{equation}
b_{\mathrm{t}}=\frac{2d}{\pi^3\phi^2} \sum_{n=1}^{\infty} \frac{\sin^2\left(n\pi\phi\right)}{n^3}.
\label{eq:bt}
\end{equation}
The Nusselt number $\overline{\mathrm{Nu}}'_{\mathrm{E}}$, where the subscript ``E'' indicates a result from \citet{Enright-14}, then follows from the definition of the bulk temperature and it does not capture the error term as in the present analysis. Consequently, the relationship between the present result and that 
by \citet{Enright-14} is
\begin{equation}
\overline{\mathrm{Nu}}'=\overline{\mathrm{Nu}}'_{\mathrm{E}} + O\left(\epsilon^3\right).
\end{equation}

Replacing $b_{\mathrm{t}}$ in the \citet{Enright-14} analysis with its maximum value, $b_{\mathrm{t,max}}$, to define the apparent temperature jump along the composite interface in terms of the temperature of the center of the ridge,  (\ref{eq:bt}) becomes
\begin{equation}
T_{\mathrm{sl}}\left(x=d\right) - \bar{T}_{\mathrm{c}} = -b_{\mathrm{t,max}} \left.\frac{\partial \bar{T}}{\partial n}\right|_{\mathrm{c}}.
\label{eq:btmax}
\end{equation}
It follows from the \citet{Mikic-57} solution that
\begin{equation}
b_{\mathrm{t,max}}=\frac{2d}{\pi^2\phi} \sum_{n=1}^{\infty} \frac{\sin\left(n\pi\phi\right)}{n^2}.
\end{equation}
Then, the \citet{Enright-14} approach yields
\begin{align}
\mathrm{Nu}_{\mathrm{min,E}} &= 140\left(1+3\epsilon \lambda\right)^2/\left\{
17\phi+ \left[84\lambda\phi + \frac{70}{\pi^2}\sum_{n=1}^{\infty}\frac{\sin\left(n\pi\phi\right)}{n^2}\right]\epsilon \right. \nonumber \\
&\left. +\left[
105\lambda^2\phi +\frac{420\lambda}{\pi^2}\sum_{n=1}^{\infty}\frac{\sin\left(n\pi\phi\right)}{n^2}
\right]\epsilon^2
 +\left[ \frac{630\lambda^2}{\pi^2}\sum_{n=1}^{\infty}\frac{\sin\left(n\pi\phi\right)}{n^2}\right]\epsilon^3\right\}. \label{eq:numinE}
\end{align}
Upon rearrangement this result may be shown to be identical to our own result as per  (\ref{eq:numinKp}), except that it does not quantify the error. 
More generally, the thermal slip length in the \citet{Enright-14} formulation may be based upon any 
location along the ridge and thus the variation of the local Nusselt number along it determined. Then, the local Nusselt number
$\overline{\mathrm{Nu}}$ follows. We compare the value of 
$\overline{\mathrm{Nu}}'_{\mathrm{E}}$ (or, equivalently, $\overline{\mathrm{Nu}}'$ from 
this study) to its exact value as per the \citet{Kirk-17} study in Fig.~\ref{fig:nupP}.
Of course, the agreement only breaks down at sufficiently large values of $\epsilon$.

Only the aforementioned exact results by \citet{Kirk-17} are valid for arbitrary $\epsilon$. Also, by taking the limit of the dual-series equations as $\epsilon \rightarrow 0$, and the same limit in the temperature problem, an expression for 
$\overline{\mathrm{Nu}}$ with an error term of $O\left(\epsilon^2\right)$ was found by \citet{Kirk-17}. It breaks
down for sufficiently small solid fraction. We summarize the Nusselt number developed in this study in Table~\ref{tab:ps}.
\begin{table}
\begin{center}
\caption{Nusselt number results for parallel ridges.}\vspace{0.1in}
\begin{tabular}{clcc}
\hline
Nusselt Number  &Description &Expression  &Error  \\
\hline
$\mathrm{Nu}$ &local & \eqref{eq:nuloc} &$O\left(\epsilon^3\right)$\\
\hline
$\mathrm{Nu_{min}}$&minimum& \eqref{eq:numinKp}&$O\left(\epsilon^3\right)$  \\ \hline
$\overline{\mathrm{Nu}}$&mean of $\mathrm{Nu}$ & \eqref{eq:nupf} &$O\left(\epsilon^3\right)$  \\\hline
$\overline{\mathrm{Nu}}'$&based on $\bar{h}'= \phi q''_{\mathrm{sl}}/(\overline{T_{\mathrm{sl}}-T_{\mathrm{m}}})$
& \eqref{eq:nupp}&$O\left(\epsilon^3\right)$  \\
\hline
\end{tabular}
\label{tab:ps}
\end{center}
\end{table}

\subsubsection{Comparisons}
\begin{figure}
\begin{centering}
\includegraphics[width=9cm]{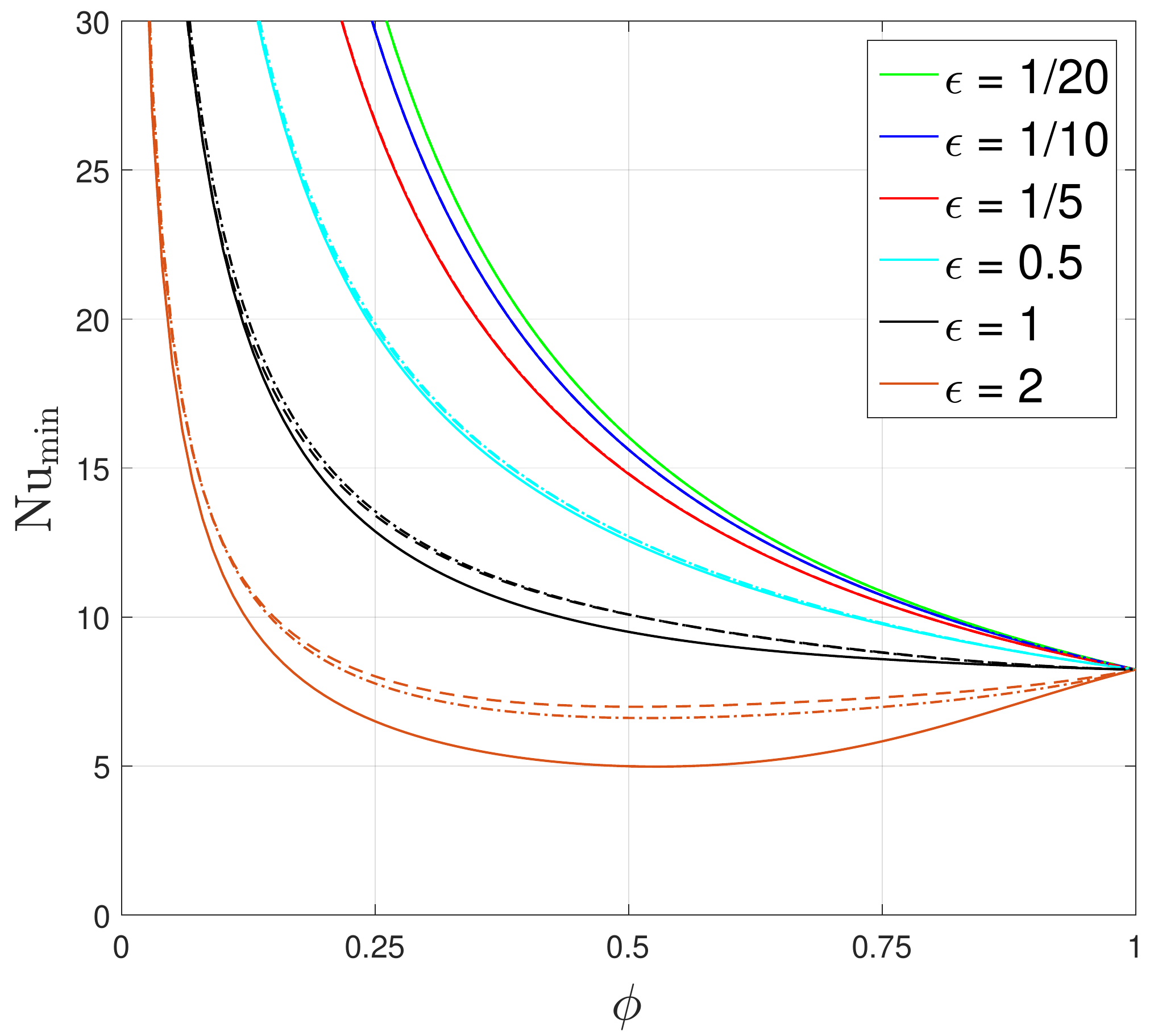}
\par\end{centering}
\caption{${\mathrm{Nu}}_{\mathrm{min}}$ vs.~$\phi$ for selected values of $\epsilon$ from present result (or, equivalently,
that which follows from the approach of \citet{Enright-14}) per  (\ref{eq:numinKp}) per dashed curves, \citet{Maynes-14} expression for composite interface temperature per
 (\ref{eq:tslM}) per dot-dash curves and
exact solution by \citet{Kirk-17} per solid curves.}
\label{fig:NuminPa}
\end{figure}

\begin{figure}
\begin{centering}
\includegraphics[width=9cm]{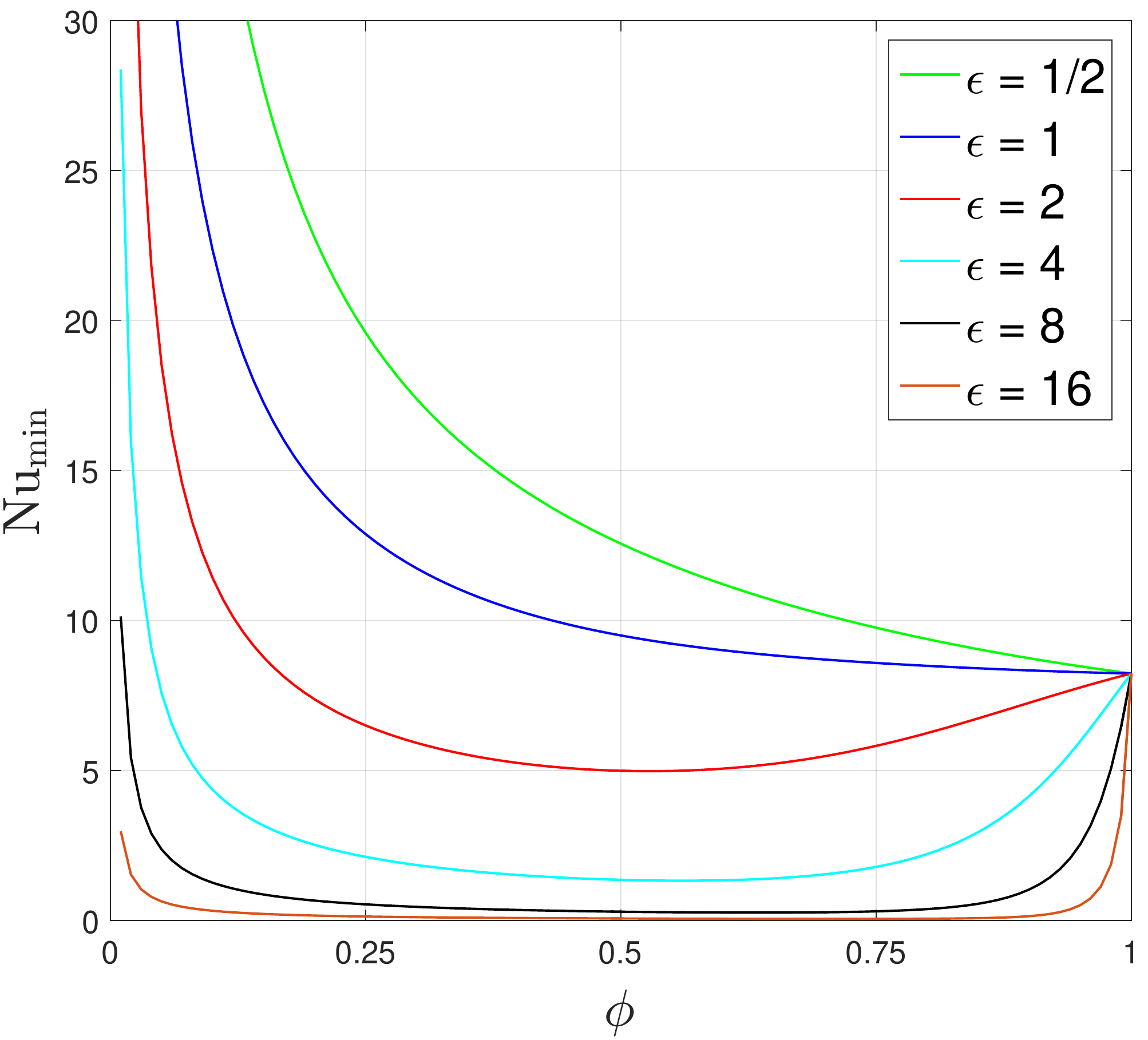}
\par\end{centering}
\caption{${\mathrm{Nu_{min}}}$ vs.~$\phi$ for selected values of relatively large $\epsilon$  calculated using the (exact) method of \citet{Kirk-17}.}
\label{fig:numinlarge}
\end{figure}

$\mathrm{Nu}_{\mathrm{min}}$ vs.~$\phi$ for selected values of $\epsilon$ based on the present result and the aforementioned studies is shown in Fig.~\ref{fig:NuminPa}. As expected, it asymptotes to $\infty$ and 140/17 as $\phi$ approaches 0 and 1, respectively. When the domain is
square ($\epsilon = 1$), our solution and the \citet{Maynes-14} one overpredict $\mathrm{Nu_{min}}$ by a maximum amount of 21.6\% and 21.7\%, respectively, at 
$\phi$ = 0.64.
When $\epsilon = 2$, a local minimum in $\mathrm{Nu}_{\mathrm{min}}$ occurs at about $\phi = 1/2$, presumably because the flow velocity is rather low above the center of the ridge relative to above the triple contact line (and meniscus). Using only the exact solution from \citet{Kirk-17}, in 
Fig.~\ref{fig:numinlarge}, we show the same results up to $\epsilon = 16$. $\mathrm{Nu}_{\mathrm{min}}$ becomes very low as $\epsilon$ is sufficiently increased, except as $\phi \rightarrow 0$ or $\phi \rightarrow 1$, because most of the flow in the domain is between opposing menisci rather than
between opposing ridges.

\begin{figure}
\begin{centering}
\includegraphics[width=9cm]{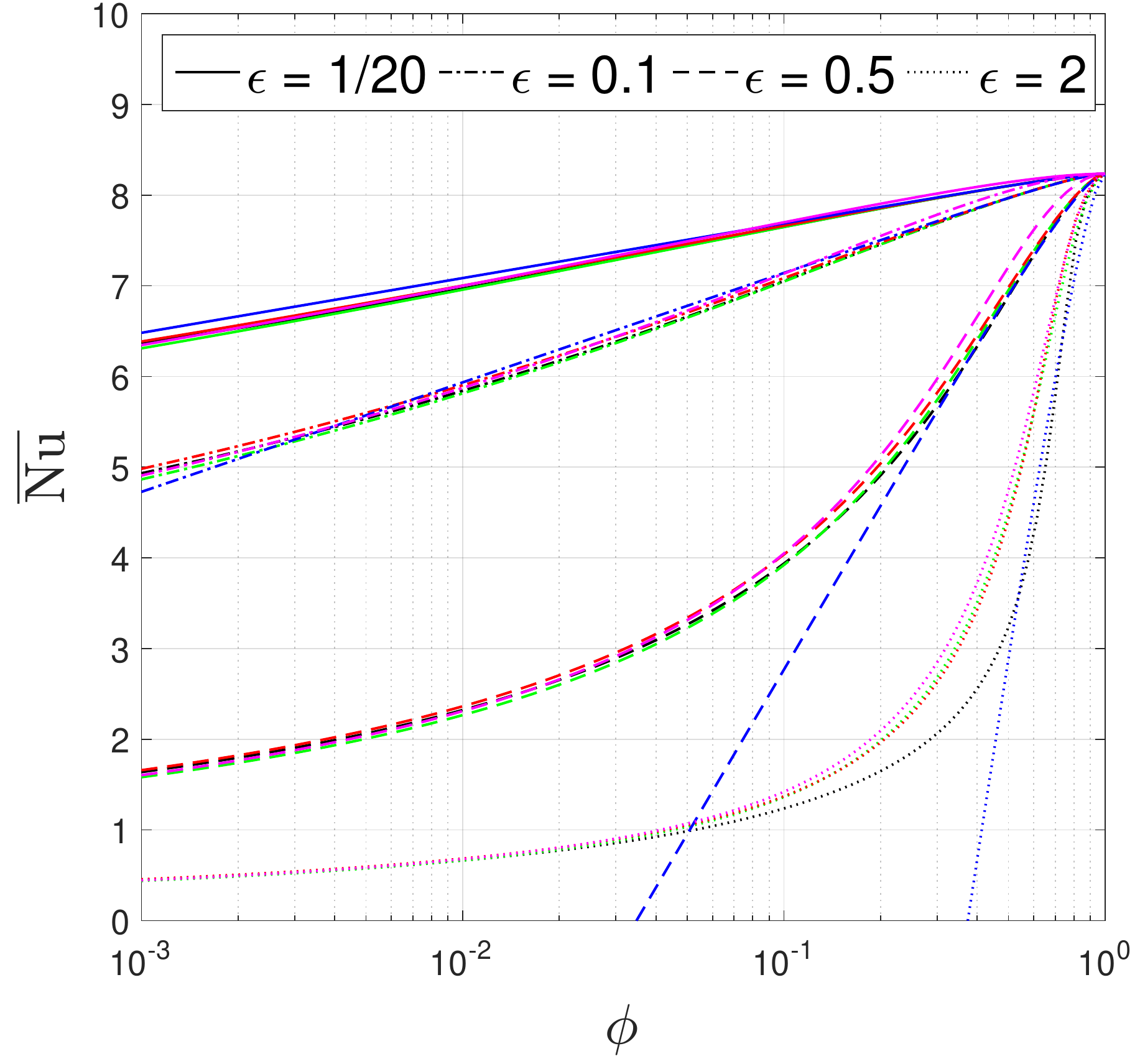}
\par\end{centering}
\caption{Semilog plot of $\overline{\mathrm{Nu}}$ vs.~$\phi$ for selected values of $\epsilon$  from present result per  (\ref{eq:nupf}) per green curves, simplified form of present result per  (\ref{eq:nupfs}) per magenta curves, numerically-evaluated integral from 
\citet{Maynes-14} per red
curves, asymptotic expression by \citet{Kirk-17} per blue curves and exact solution by \citet{Kirk-17} 
per black curves.}
\label{fig:nu1R}
\end{figure}

\begin{figure}
\begin{centering}
\includegraphics[width=9cm]{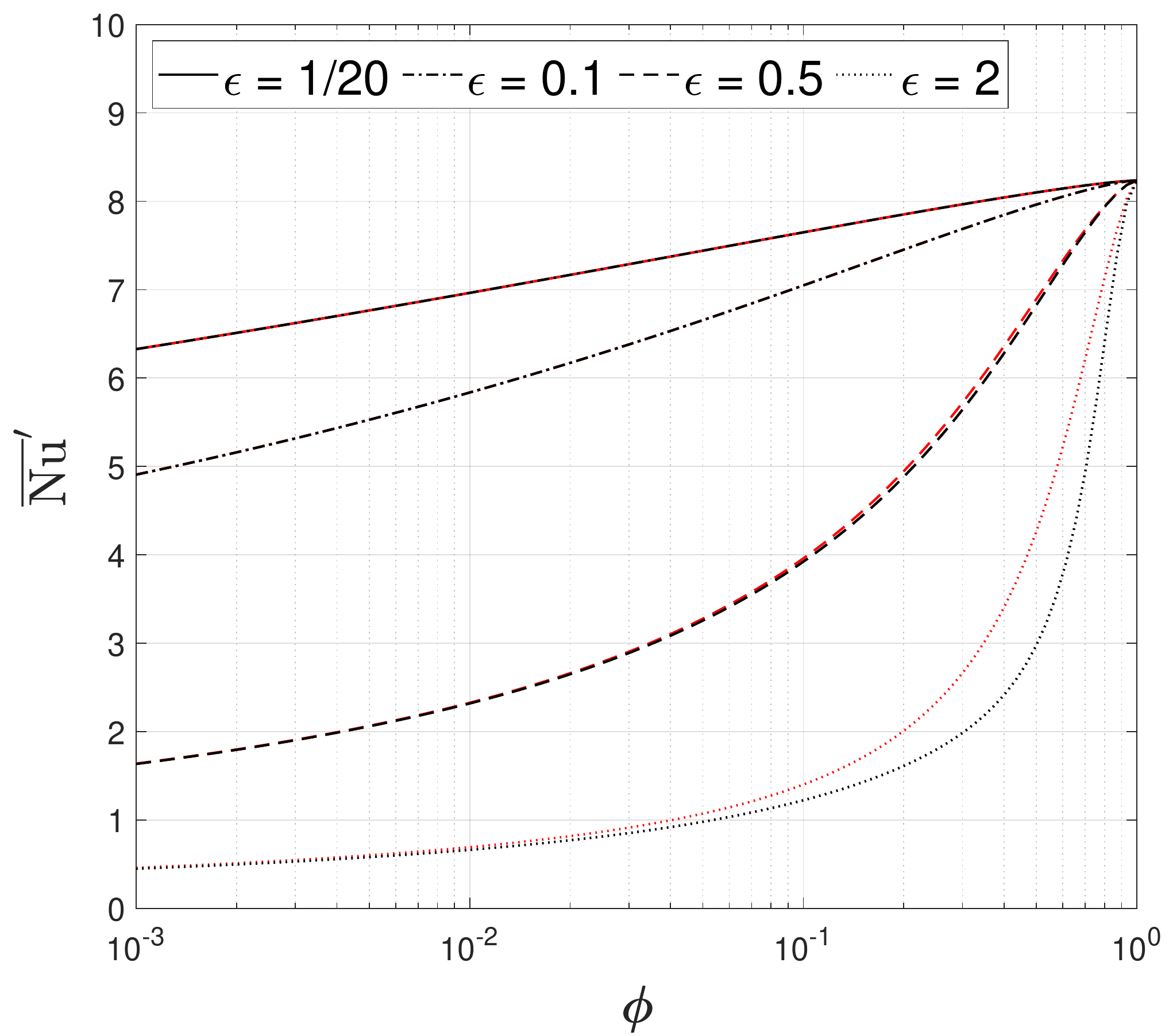}
\par\end{centering}
\caption{$\overline{\mathrm{Nu}}'$ vs.~$\phi$ for $\epsilon$ = 1/20, 0.1, 0.5 and 2 from Enright et
al., or, equivalently, the present study,  \eqref{eq:nupp}, per red curves
 and exact solution by \citet{Kirk-17} per black curves.}
\label{fig:nupP}
\end{figure}

A semilog plot of $\overline{\mathrm{Nu}}$ vs.~$\phi$ for selected values of $\epsilon$ based on our results and those from the aforementioned studies is shown in Fig.~\ref{fig:nu1R}. When $\epsilon = 0.5$, a substantive value, the maximum discrepancies between $\overline{\mathrm{Nu}}$ from \citet{Maynes-14}, the present result with
an error term of $O\left(\epsilon^3\right)$,  (\ref{eq:nupf}), and that with an error term of 
$O\left(\epsilon\right)$,  \eqref{eq:nupfs}, compared to the exact solution are only 2.6\%, 1.2\% and 5.2\%, respectively. When $\epsilon$ is  increased to 2, they become 36.6\%, 38.7\% and 46.5\%, respectively, occuring near a solid fraction of  0.4. An extensive discussion of the physics governing the behavior of curves in Fig.~\ref{fig:nu1R}, except that given by  (\ref{eq:nupf}), is given by \citet{Kirk-17}. The behavior of the latter is analogous to that of the asymptotic result in \citet{Kirk-17}, except that the error is $O\left(\epsilon^3\right)$ rather than $O\left(\epsilon^2\right)$ and it does not break down at small $\phi$. Finally, Fig.~\ref{fig:nupP} shows the comparison between the Enright et al. result (equivalently,  \eqref{eq:nupp}) and the exact solution, which is rather good even up to $\epsilon = 2$.

\subsubsection{Other Relevant Studies}
Beyond the problem considered here, \citet{Kirk-17} accounted for meniscus curvature, assuming small deflection from flat, by using a boundary perturbation. Also, the numerical results by \citet{Game-18} consider arbitrary meniscus curvature, so long as it's a circular arc. Finally, \citet{Karamanis-18} solved the extended Graetz-Nusselt problem in the presence of viscous dissipation and uniform volumetric heat generation for a flat meniscus assuming isothermal, parallel ridges. Nusselt number results for transverse ridges are provided below. Those for (square) pillars
using the \citet{Enright-14} approach are compared to numerical results in that study. 
Finally, very recently, \citet{Sharma-20} provided numerical results for square, triangular
and herringbone pillars in regular and staggered arrays.

\subsubsection{Alternate Form of Inner Solution and Corresponding Spreading Resistances}

Since the heat flux is discontinuous along the composite interface ($Y = 0$), the series in  (\ref{eq:mikic}) converges non-uniformly in the domain ($Y > 0$) and requires many terms to accurately represent the
temperature and heat flux profiles near the composite interface. Therefore, we ``sum the series'' to obtain the exact solution in closed form. The dimensionless temperature field is a harmonic function, which can be written in terms of the imaginary part of a complex potential as per
\begin{equation}
\theta=\operatorname{Im} \left[f\left(\Theta_{\parallel}\right)\right], 
\end{equation}
where, recall, $\Theta_{\parallel} = X + \mathrm{i}Y$. Judicious use of Euler's formula (as
in \citet{Bazant-04}, Sec.~V.G.) shows that
\begin{equation}
f\left(\Theta_{\parallel}\right)=-\epsilon
\left\{ \phi \Theta_{\parallel}+\frac{\mathrm{Li}_{2}\left(M_{+}\right)-\mathrm{Li}_{2}\left(M_{-}\right)}{\pi^{2}}\right\} + \mathrm{i}\phi\left[\frac{1}{2}+\frac{1/12+D(\epsilon)}{2/3+2\epsilon \lambda}\right]+O\left(\epsilon^{3}\right),
\label{eq:inth}
\end{equation}
where the polylogarithm (special) function is 
\begin{equation}
\mathrm{Li}_{s}\left(M\right)=\sum_{n=1}^{\infty}\frac{M^{n}}{n^{s}},~~~~\left|M\right|<1 \label{eq:DiLogarithm}
\end{equation}
and
\begin{equation}
M_{\pm}=e^{\mathrm{i}\pi\left(\Theta_{\parallel}\pm\delta\right)}.
\end{equation}
Appealingly, in that no special functions are required, the dimensionless heat flux vector follows as
\begin{equation}
-\nabla \theta =-\left(\frac{\partial \theta}{\partial X} + \mathrm{i}\frac{\partial \theta}{\partial Y}\right) = 
\frac{1}{\mathrm{i}}\overline{f'\left(\Theta_{\parallel}\right)}, 
\end{equation}
where, using the relation that $\mathrm{d}\mathrm{Li}_2\left( M_{\pm}\right)/\mathrm{d}M_{\pm} = -\ln\left(1-M_{\pm}\right)/M_{\pm}$,
\begin{equation}
f'\left(\Theta_{\parallel}\right) =-\epsilon \phi +\frac{\mathrm{i}\epsilon}{\pi}\left[\ln\left(1-M_+\right)-\ln\left(1-M_-\right)\right].
\end{equation}

The preceding result is also useful in the context of spreading resistance, $R_{\mathrm{sp}}$, which we define conventionally. Hence, it's the increase in the temperature difference driving heat transfer, i.e., the mean heat source temperature minus the far-field temperature, beyond that of the corresponding one-dimensional problem, divided by the heat rate. Thus, the (dimensional) spreading resistance based upon the heat rate per unit depth of the domain, i.e., in units of W/(m$\cdot$K), is
\begin{equation}
R'_{\mathrm{sp}} = \frac{\int_a^d T\left(x,0\right)\mathrm{d}x/\left(d-a\right) -\lim_{y\rightarrow \infty}T}{q''_{\mathrm{sl}}\left(d-a\right)} - \frac{T_{\mathrm{1d}}\left(0\right) -\lim_{y\rightarrow \infty}T_{\mathrm{1d}}}{q''_{\mathrm{sl}}\left(d-a\right)},
\end{equation}
where $T_{\mathrm{1d}}\left(y\right)$ is, to within an additive constant, $-q''_{\mathrm{sl}}\left(d-a\right)y/\left(kd\right)$.
Non-dimensionalizing spreading resistance by $1/k$, the well-known result by \citet{Mikic-57} is
\begin{equation}
\tilde{R}'_{\mathrm{sp}} = \frac{2}{\pi^3\phi^2} \sum_{n=1}^{\infty}
\frac{\sin^2\left(n\pi\phi\right)}{n^3} 
\label{eq:clm}
\end{equation}
or, albeit not previously realized, in closed form
\begin{equation}
\tilde{R}'_{\mathrm{sp}} = \frac{1}{\pi^3\phi^2}\mathrm{Re}\left[\mathrm{Li}_3\left(1\right) - \mathrm{Li}_3\left(\mathrm{e}^{\mathrm{i}2\pi\phi}\right)
\right].
\end{equation}
This result is an exact result as the spreading resistance problem is one of pure diffusion in a semi-infinite domain. It's twice the value reported by 
\citet{Mikic-57} because the width of his domain corresponds to one ridge pitch (i.e., the full width of the heat source) and ours is half of that. (The spreading resistance based on the far-field heat flux, $R''_{\mathrm{sp}}$, does not depend on the domain width and, non-dimensionalizing it by $\left(d-a\right)/k$, it follows that $\tilde{R}''_{\mathrm{sp}}= \tilde{R}'_{\mathrm{sp}}$.) More conservatively, the corresponding dimensionless spreading resistance based on the maximum source temperature, $\tilde{R}'_{\mathrm{sp,max}}$ (or, equivalently, $\tilde{R}''_{\mathrm{sp,max}}$), is
\begin{equation}
\tilde{R}'_{\mathrm{sp,max}} = \frac{2}{\pi^2\phi}\sum_{n=1}^{\infty}\frac{\sin\left(n\pi\phi\right)}{n^2}
\end{equation}
or, in closed form,
\begin{equation}
\tilde{R}'_{\mathrm{sp,max}} =\frac{2}{\pi^2\phi}\mathrm{Im}\left[\mathrm{Li}_2\left(\mathrm{e}^{i\pi\phi}\right)\right].
\end{equation}

In the context of the asymptotics  performed here, if a constant heat flux is applied along the top of a finite rather than infinite height domain, the error in spreading resistance is exponentially small as $\epsilon \rightarrow 0$. For example,  (\ref{eq:clm}) would become 
\begin{equation}
\tilde{R}'_{\mathrm{sp,a}} \sim \frac{2}{\pi^3\phi^2} \sum_{n=1}^{\infty} 
\frac{\sin^2\left(n\pi\phi\right)}{n^3} + \mathrm{e.s.t.},~~~\epsilon \rightarrow 0,
\end{equation}
where the subscript ``sp,a'' signifies this is an asymptotic limit. Clearly, the foregoing results may
be used to eliminate some of the infinite summations in the results that we provided. Finally, we note that 
\citet{Hodes-18} discusses analogies between the slip length for a linear-shear flow 
and spreading and contact resistances.

\section{Transverse Ridges}
\begin{figure}
\begin{centering}
\includegraphics[width=10cm]{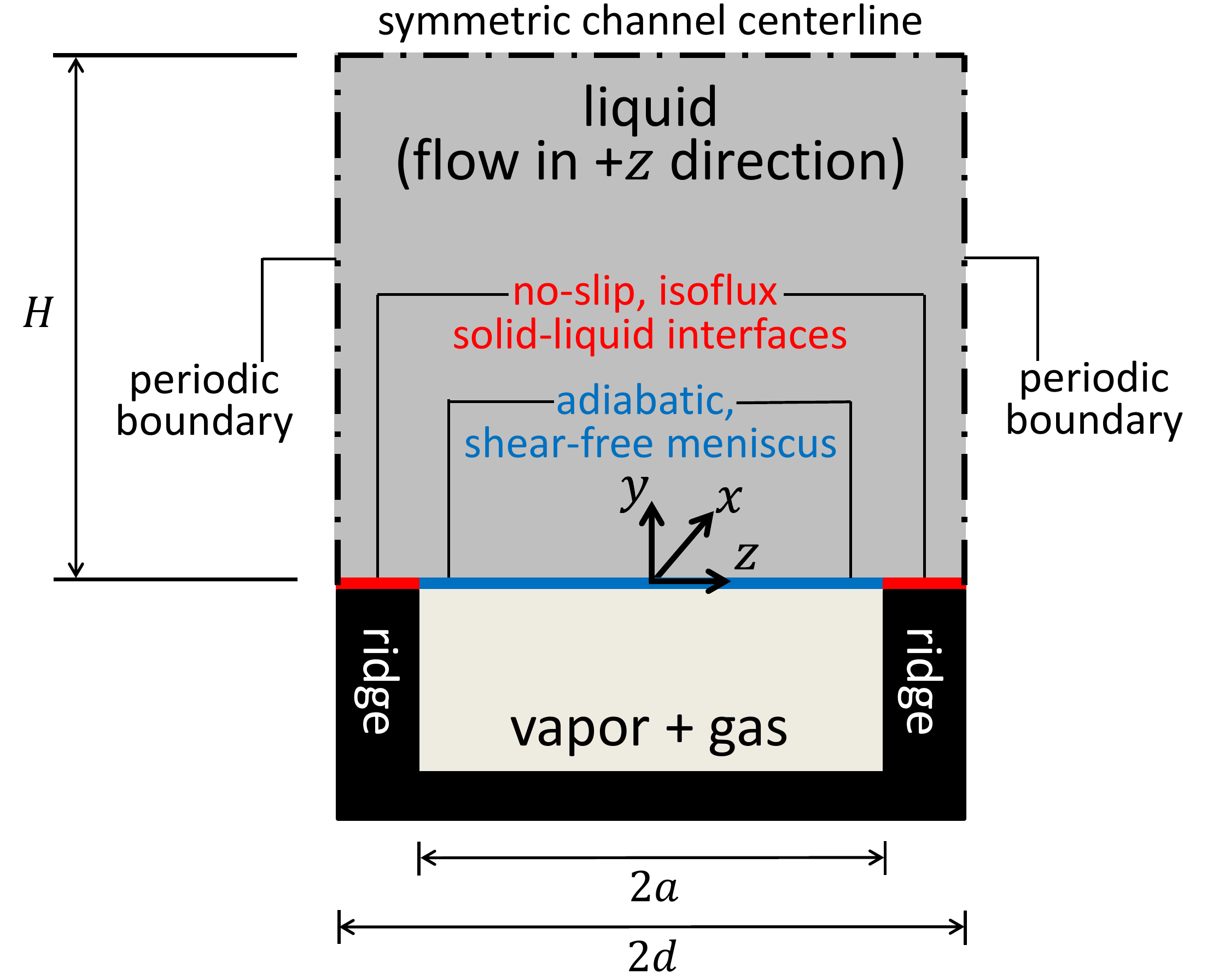}
\par\end{centering}
\protect\caption{Schematic of transverse-ridge problem with the (planar) domain shown in gray. 
Imposition of a (negative) linear component of the streamwise-pressure
gradient ($\beta$) drives the flow in the $z$-direction.\label{fig:dt}}
\end{figure}

A schematic of the (dimensional) geometry of the transverse-ridge problem is shown in Fig.~\ref{fig:dt}
(left). The periodically fully-developed, bidirectional flow in the streamwise ($z$) and transverse ($y$) directions is driven by prescribing the linear component of the streamwise pressure gradient as a (negative) constant ($\beta$). The velocity field is periodic in the $z$-direction, whereas the pressure and temperature fields have both linear and periodic components as per the class of diabatic flows treated by \citet{Patankar-77}. The ridge and channel geometry are 
described the smae as in Section~\ref{sec:parallel} (Fig.~\ref{fig:Parallel_Domain-and-boundary}), but now with the ridges
aligned with the $x$-axis rather than the $z$-axis.
 Restricting our 
attention to  the Stokes flow limit, the domain length reduces to $d$ due to mirror symmetry about the center of the meniscus in the periodic components of the hydrodynamic problem. The same holds in the low P{\'e}clet number limit for the thermal problem
 due to symmetry about the center of the meniscus.

\subsection{Hydrodynamic Problem}
Following \citet{Patankar-77}, we decompose the pressure field into
\begin{equation}
p=\beta z+p_{\mathrm{p}}\left(y,z\right),
\end{equation}
where $p_{\mathrm{p}}\left(y,z\right)$, its periodic component, repeats itself in the streamwise direction with a period of one pitch ($2d$). We nondimensionalize velocities by $-\beta H^2/\left(2\mu\right)$, $p_{\mathrm{p}}$ by $\beta H$
and lengths as in Section~\ref{sub:Hydrodynamic}. Working in terms of $Z=\tilde{z}/\epsilon$ such that the extents of the domain are independent of the small parameter, the dimensionless 
continuity, streamwise- and transverse-momentum equations become
\begin{align}
\frac{\partial\tilde{v}}{\partial\tilde{y}}+\frac{1}{\epsilon}\frac{\partial\tilde{w}}{\partial{Z}}&=0 \label{eq:tcon} \\
\mathrm{Re}\left(\tilde{v}\frac{\partial\tilde{v}}{\partial\tilde{y}}+\frac{\tilde{w}}{\epsilon}\frac{\partial\tilde{v}}{\partial Z}\right)&=2\frac{\partial\tilde{p}_{\mathrm{p}}}{\partial\tilde{y}}+\left(\frac{\partial^{2}\tilde{v}}{\partial\tilde{y}^{2}}+\frac{1}{\epsilon^2}\frac{\partial^{2}\tilde{v}}{\partial{Z}^{2}}\right) \label{eq:ttm}\\
\mathrm{Re}\left(\tilde{v}\frac{\partial\tilde{w}}{\partial\tilde{y}}+\frac{\tilde{w}}{\epsilon}\frac{\partial\tilde{w}}{\partial Z}\right)&=2+\frac{2}{\epsilon}\frac{\partial\tilde{p}_{\mathrm{p}}}{\partial{Z}}+\left(\frac{\partial^{2}\tilde{w}}{\partial\tilde{y}^{2}}+\frac{1}{\epsilon^2}\frac{\partial^{2}\tilde{w}}{\partial{Z}^{2}}\right), \label{eq:tsm}
\end{align}
respectively, where $\tilde{v}$ is the transverse velocity and the Reynolds number is
\begin{equation}
\mathrm{Re}=-\frac{\rho \beta H^3}{2 \mu^2}.
\end{equation}
The impermeability and shear-free boundary conditions along the meniscus, impermeability and no-slip ones along the solid-liquid interfaces and mirror symmetry one about the centerline of the channel manifest themselves as
\begin{align}
\tilde{v}&=\frac{\partial\tilde{w}}{\partial\tilde{y}}=0\,\,\,\mathrm{at}\,\,\,\tilde{y}=0\,\,\,\mathrm{for}\,\,\,0<\vert{Z}\vert<\delta 
\label{eq:bcv}\\
\tilde{v}&=\tilde{w}=0\,\,\,\mathrm{at}\,\,\,\tilde{y}=0\,\,\,\mathrm{for}\,\,\,\delta<\vert{Z}\vert<1\\
\tilde{v}&=\frac{\partial\tilde{w}}{\partial\tilde{y}}=\partial p_{\mathrm{p}}/\partial \tilde{y}=0\,\,\,\mathrm{at}\,\,\,\tilde{y}=1\,\,\,\mathrm{for}\,\,\,0<\vert{Z}\vert<1,
\end{align}
respectively. The (streamwise) periodicity boundary condition is
\begin{equation}
\chi(\tilde{y},-1)=\chi(\tilde{y},1),\,\,\,\mathrm{where}\,\,\,\chi=\tilde{v},\tilde{w},\tilde{p}_{\mathrm{p}},\frac{\partial\tilde{v}}{\partial Z},\frac{\partial\tilde{w}}{\partial Z}.
\label{eq:ph}
\end{equation}
As an ansatz, we assume that $\tilde{v}$, $\tilde{w}$ and $\tilde{p}_{\mathrm{p}}$ are $O\left(1\right)$ as
$\epsilon \rightarrow 0$ and thus asymptotically expand them as 
\begin{align}
\tilde{v}&\sim\sum_{n=0}^{\infty}\epsilon^{n}\tilde{v}_{n} + \mathrm{e.s.t.}
\label{eq:tve}\\
\tilde{w}&\sim \sum_{n=0}^{\infty}\epsilon^{n}\tilde{w}_{n} \label{eq:trans_w_expand}+ \mathrm{e.s.t.}\\
\tilde{p}_{\mathrm{p}}&\sim \sum_{n=0}^{\infty}\epsilon^{n}\tilde{p}_{\mathrm{p,}n}+ \mathrm{e.s.t.}\label{eq:tvp},
\end{align}
where $\tilde{v}_{n}$, $\tilde{w}_{n}$ and $\tilde{p}_{\mathrm{p},n}$ are O(1) for
$n\geq 0$.

\subsubsection{Outer Region}
The outer region is where $Z=O(1)$ and $\tilde{y}=\mathrm{ord}(1)$ as $\epsilon\rightarrow0$.  Substituting 
 (\ref{eq:tve}-\ref{eq:tvp}) into  (\ref{eq:tcon}-\ref{eq:tsm}) yields, at the algebraic orders of $\epsilon$,
\begin{align} 
O\left(\epsilon^{-2}\right):~~&\frac{\partial^{2} \tilde{v}_{0}}{\partial Z^2}=0 \label{eq:vl}\\
&\frac{\partial^{2} \tilde{w}_{0}}{\partial Z^2}=0 \label{eq:wl}
\end{align}
\begin{align}
O\left(\epsilon^{-1}\right):~~&\frac{\partial \tilde{w}_{0}}{\partial Z}=0 \\ 
&\mathrm{Re}\left(\tilde{w}_{0}\frac{\partial \tilde{v}_{0}}{\partial Z}\right)=\frac{\partial^{2}\tilde{v}_{1}}{\partial Z^{2}} \label{eq:vmm1}\\
&\mathrm{Re}\left(\tilde{w}_{0}\frac{\partial \tilde{w}_{0}}{\partial Z}\right)=2\frac{\partial \tilde{p}_{\mathrm{p,}0}}{\partial Z}+\frac{\partial^{2}\tilde{w}_{1}}{\partial Z^{2}} \label{eq:wmm1}
\end{align}
\begin{align}
O\left(\epsilon^{0}\right):~~&\frac{\partial \tilde{v}_{0}}{\partial\tilde{y}}+\frac{\partial \tilde{w}_{1}}{\partial Z}=0 \label{eq:zoc} \\ 
&\mathrm{Re}\left(\tilde{v}_{0}\frac{\partial\tilde{v}_{0}}{\partial\tilde{y}}+\tilde{w}_{0}\frac{\partial \tilde{v}_{1}}{\partial Z}+\tilde{w}_{1}\frac{\partial\tilde{v}_{0}}{\partial Z}\right)=2\frac{\partial\tilde{p}_{\mathrm{p,}0}}{\partial\tilde{y}}+\frac{\partial^{2}\tilde{v}_{0}}{\partial\tilde{y}^{2}}+\frac{\partial^{2}\tilde{v}_{2}}{\partial Z^{2}} \label{eq:v_mom_zero}\\
&\mathrm{Re}\left(\tilde{v}_{0}\frac{\partial\tilde{w}_{0}}{\partial\tilde{y}}+\tilde{w}_{0}\frac{\partial \tilde{w}_{1}}{\partial Z}+\tilde{w}_{1}\frac{\partial\tilde{w}_{0}}{\partial Z}\right)=2+2\frac{\partial\tilde{p}_{\mathrm{p,}1}}{\partial Z}+\frac{\partial^{2}\tilde{w}_{0}}{\partial\tilde{y}^{2}}+\frac{\partial^{2}\tilde{w}_{2}}{\partial Z^{2}} \label{eq:w_mom_zero}
\end{align}
\begin{align}
O\left(\epsilon^{n}\right):~~&\frac{\partial\tilde{v}_{n}}{\partial\tilde{y}}+\frac{\partial\tilde{w}_{n+1}}{\partial Z}=0~\mathrm{for}~n\geq 0\label{eq:cf}\\
&\mathrm{Re}\left(\sum_{m=0}^n \tilde{v}_m \frac{\partial \tilde{v}_{n-m}}{\partial \tilde{y}}
+ \sum_{m=0}^{n+1} \tilde{w}_m \frac{\partial \tilde{v}_{n+1-m}}{\partial Z}
\right)=2\frac{\partial\tilde{p}_{\mathrm{p,}n}}{\partial\tilde{y}}+\frac{\partial^{2}\tilde{v}_{n}}
{\partial\tilde{y}^2}+\frac{\partial^{2}\tilde{v}_{n+2}}{\partial Z^{2}} \label{eq:vm1}
~\mathrm{for}
~n\geq 1\\	
&\mathrm{Re}\left(\sum_{m=0}^n \tilde{v}_m \frac{\partial \tilde{w}_{n-m}}{\partial \tilde{y}}
+ \sum_{m=0}^{n+1} \tilde{w}_m \frac{\partial \tilde{w}_{n+1-m}}{\partial Z}\right)=
2\frac{\partial\tilde{p}_{\mathrm{p,}n+1}}{\partial Z}+\frac{\partial^{2}\tilde{w}_{n}}{\partial\tilde{y}^2}+\frac{\partial^{2}\tilde{w}_{n+2}}{\partial Z^{2}}~\mathrm{for}
~n\geq 1. \label{eq:wm1}
\end{align}
The boundary conditions as per  (\ref{eq:bcv}-\ref{eq:ph}) apply at all orders.

Integrating the leading-order momentum equations, i.e.,  (\ref{eq:vl}) and~(\ref{eq:wl}), twice and applying periodicity shows that  
$\tilde{v}_0 = \tilde{v}_0\left(\tilde{y}\right)$ and $\tilde{w}_0 = \tilde{w}_0\left(\tilde{y}\right)$. Then, the leading-order continuity equation,  (\ref{eq:zoc}), becomes $\partial\tilde{w}_{1}/\partial Z = -\mathrm{d}\tilde{v}_0/\mathrm{d}\tilde{y}.$
Integrating it with respect to $Z$ and applying periodicity on $\tilde{w}_1$ shows that 
$\mathrm{d}\tilde{v}_{0}/\mathrm{d}\tilde{y}=0$ and thus from the mirror symmetry condition at $\tilde{y}=1$ that $\tilde{v}_{0}=0$, and $\tilde{w}_{1} = \tilde{w}_{1}\left(\tilde{y}\right)$. Relatedly,  (\ref{eq:vmm1}) and (\ref{eq:wmm1}) collapse to
$\partial^2 \tilde{v}_1/\partial Z^2 = 0$ and $\partial \tilde{p}_{\mathrm{p,}0}/\partial Z =0$, respectively. By implication, $\tilde{v}_{1}=\tilde{v}_{1}\left(\tilde{y}\right)$ and $\tilde{p}_{\mathrm{p,}0}=\tilde{p}_{\mathrm{p,}0}\left(\tilde{y}\right)$. In turn, 
 (\ref{eq:cf}) along with the boundary conditions implies that $\tilde{v}_{1}=0$ and $\tilde{w}_{2} = \tilde{w}_{2}\left(\tilde{y}\right)$. Equations (\ref{eq:v_mom_zero}) and (\ref{eq:w_mom_zero}) then reduce to $\partial^{2}\tilde{v}_{2}/\partial Z^{2}=-2\,\mathrm{d}\tilde{p}_{\mathrm{p,}0}/\mathrm{d}\tilde{y}$
and $ \partial\tilde{p}_{\mathrm{p},1}/\partial Z = -\mathrm{d}^{2}\tilde{w}_{0}/\mathrm{d}\tilde{y}^{2}/2-1 $, respectively. Integrating the former with respect to $Z$ and applying periodicity shows that $\mathrm{d}\tilde{p}_{\mathrm{p,}0}/\mathrm{d}\tilde{y} = 0$ and thus $\tilde{p}_{\mathrm{p,}0}$ is a constant ($E_0$), and doing so again that 
$\tilde{v}_{2}=\tilde{v}_{2}\left(\tilde{y}\right)$. Integrating the latter with respect to $Z$ and applying periodicity shows that $\tilde{p}_{\mathrm{p,}1}=\tilde{p}_{\mathrm{p,}1}\left(\tilde{y}\right)$ and $\mathrm{d}^{2}\tilde{w}_{0}/\mathrm{d}\tilde{y}^{2}=-2$. Therefore, to satisfy the mirror symmetry condition, 
\begin{equation}
\tilde{w}_{0}=-\tilde{y}^{2}+2\tilde{y}+G_{0},
\end{equation}
where $G_0$ is a constant.
In turn, the $O\left(\epsilon^2\right)$ continuity equation (not shown) along with the boundary conditions 
implies that $\tilde{v}_{2}=0$ and $\tilde{w}_{3} = \tilde{w}_{3}\left(\tilde{y}\right)$. Then,  (\ref{eq:vm1}) and (\ref{eq:wm1})  reduce to $\partial^{2}\tilde{v}_{3}/\partial Z^{2}=-2\,\mathrm{d}\tilde{p}_{\mathrm{p,}1}/\mathrm{d}\tilde{y}$
and $\partial\tilde{p}_{\mathrm{p},2}/\partial Z=-\mathrm{d}^{2}\tilde{w}_{1}/\mathrm{d}\tilde{y}^{2}/2\,$, respectively. 
The  former leads to $\tilde{p}_{\mathrm{p,}1} = E_1$ and $\tilde{v}_3=\tilde{v}_3\left(\tilde{y}\right)$. The latter leads to $\tilde{p}_{\mathrm{p,}2}=\tilde{p}_{\mathrm{p,}2}\left(\tilde{y}\right)$ and $\tilde{w_1} = G_1$. Generalizing the foregoing process for $n \geq 2$, the $O\left(\epsilon^n\right)$ continuity equation leads to $\tilde{v}_n = 0$ and, in turn, the $O\left(\epsilon^{n-1}\right)$ 
transverse- and streamwise-momentum equations lead to 
$\tilde{p}_{\mathrm{p},n-1} = E_{n-1}$ and $\tilde{w}_{n-1} = G_{n-1}$, respectively. 
This leads to a unidirectional velocity profile to all algebraic orders in the outer region as per 
\begin{align}
\tilde{v} &\sim \mathrm{e.s.t.} \label{eq:voutert} \\
\tilde{w}&\sim -\tilde{y}^{2}+2\tilde{y}+G\left(\epsilon\right)+ \mathrm{e.s.t.},\label{eq:tov}
\end{align}
where
\begin{equation}
G\left(\epsilon\right)=\sum_{n=0}^{\infty}G_{n}\epsilon^{n}. \label{eq:Gdef}
\end{equation}
The corresponding periodic component of the pressure field is $p_{\mathrm{p}} = E\left(\epsilon\right)$, where $E\left(\epsilon\right) = \sum_{n=0}^{\infty}E_{n}\epsilon^{n}$, but it need not be resolved to find the  slip length and Nusselt number.

\subsubsection{Inner Region}
As in the case of parallel ridges, the inner
region is where $Y\sim Z=O(1)$ as $\epsilon\rightarrow0$, but, unlike in the case of parallel ridges, the flow is bidirectional. 
Using notation such that $\tilde{v}=V(Y,Z)$, $\tilde{w}=W(Y,Z)$ and $\tilde{p}_{\mathrm{p}}={P}_{\mathrm{p}}(Y,Z)$ in this region, the problem becomes
\begin{align}
\frac{\partial V}{\partial Y}+\frac{\partial W}{\partial Z}&=0\label{eq:trans_inner_cont}\\
\frac{\mathrm{Re}}{\epsilon}\left(V\frac{\partial V}{\partial Y}+W\frac{\partial V}{\partial Z}\right)&=\frac{2}{\epsilon}\frac{\partial P_{\mathrm{p}}}{\partial Y}+\frac{1}{\epsilon^{2}}\left(\frac{\partial^{2}V}{\partial Y^{2}}+\frac{\partial^{2}V}{\partial Z^{2}}\right)\label{eq:trans_inner_v_mom}\\
\frac{\mathrm{Re}}{\epsilon}\left(V\frac{\partial W}{\partial Y}+W\frac{\partial W}{\partial Z}\right)&=2+\frac{2}{\epsilon}\frac{\partial P_{\mathrm{p}}}{\partial Z}+\frac{1}{\epsilon^{2}}\left(\frac{\partial^{2}W}{\partial Y^{2}}+\frac{\partial^{2}W}{\partial Z^{2}}\right)\label{eq:trans_inner_w_mom}
\end{align}
subject to 
\begin{align}
{V}&=\frac{\partial W}{\partial Y}=0\,\,\,\mathrm{at}\,\,\,Y=0\,\,\,\mathrm{for}\,\,\,0<\vert{Z}\vert<\delta 
\label{eq:bcV}\\
V&=W=0\,\,\,\mathrm{at}\,\,\,Y=0\,\,\,\mathrm{for}\,\,\,\delta<\vert{Z}\vert<1\\
V&\sim \mathrm{e.s.t.},~W\sim-\epsilon^{2}Y^{2}+2\epsilon Y+G(\epsilon)+\mathrm{e.s.t.},~{P}_{\mathrm{p}}\sim 
E\left(\epsilon \right)+\mathrm{e.s.t.}\nonumber\\
&\mathrm{as}\,\,\, Y\rightarrow\infty~\mathrm{for}\,\,\,0<\vert{Z}\vert<1 \label{eq:tma}\\
\chi(Y,-1)&=\chi(Y,1)\,\,\,\mathrm{where}\,\,\,\chi=V,W,{P}_{\mathrm{p}},\frac{\partial V}{\partial Z},\frac{\partial W}{\partial Z},
\end{align}
where  (\ref{eq:tma}) is the Van Dyke matching condition. We expand our velocities and the periodic component of the pressure field as
\begin{align}
V&\sim \sum_{n=0}^{\infty}\epsilon^{n}V_{n}+\mathrm{e.s.t.}\\
W&\sim \sum_{n=0}^{\infty}\epsilon^{n}W_{n}+\mathrm{e.s.t.}\\
P_{\mathrm{p}}&\sim \sum_{n=0}^{\infty}\epsilon^{n} P_{\mathrm{p},n}+\mathrm{e.s.t.}
\end{align}

Substituting the expansions into the momentum equations shows that, at leading (algebraic) order,
i.e., $O\left(\epsilon^{-2}\right)$, $\nabla^{2}V_{0}=0$ and 
$\nabla^{2}W_{0}=0$.
There is no shear rate at this order as per the matching condition; consequently, $V_{0}=W_{0}=0$
(and $V$ and $W$ are $o(1)$) and thus $G_{0}=0$ (and $G\left(\epsilon\right) = o\left(1\right)$).
The momentum equations become
\begin{align}
\nabla^{2}V &= -2\epsilon\frac{\partial P_{\mathrm{p}}}{\partial Y}+O\left(\epsilon^{3}\mathrm{Re}\right) \\
\nabla^{2}W &=-2\epsilon^{2}-2\epsilon\frac{\partial P_{\mathrm{p}}}{\partial Z}+O\left(\epsilon^{3}\mathrm{Re}\right),
\end{align}
which shows that neglecting the inertial terms in them introduces $O\left(\epsilon^3\mathrm{Re}\right)$ error.

Writing
\begin{align}
V &= 2\epsilon\widehat{V}\\
W &=-\epsilon^{2}Y^{2}+2\epsilon\widehat{W}
\end{align}
we find that $\widehat{V}$ and $\widehat{W}$ satisfy 
\begin{align}
\frac{\partial \widehat{V}}{\partial Y} + \frac{\partial \widehat{W}}{\partial Z}&=0\\
\nabla^{2}\widehat{V} &= -\frac{\partial P_{\mathrm{p}}}{\partial Y}+O\left(\epsilon^{2}\mathrm{Re}\right)\\
\nabla^{2}\widehat{W} &=-\frac{\partial P_{\mathrm{p}}}{\partial Z}+O\left(\epsilon^{2}\mathrm{Re}\right)
\end{align}
subject to
\begin{align}
\widehat{V}&=\frac{\partial \widehat{W}}{\partial Y}=0\,\,\,\mathrm{at}\,\,\,Y=0\,\,\,\mathrm{for}\,\,\,0<\vert{Z}\vert<\delta \\
\widehat{V}&=\widehat{W}=0\,\,\,\mathrm{at}\,\,\,Y=0\,\,\,\mathrm{for}\,\,\,\delta<\vert{Z}\vert<1\\
\widehat{V}&\sim \mathrm{e.s.t},~\widehat{W}\sim 
Y+\frac{G(\epsilon)}{2\epsilon}+\mathrm{e.s.t.},~{P}_{\mathrm{p}}\sim E\left(\epsilon\right)+\mathrm{e.s.t.}~\mathrm{as}\,\,\, Y\rightarrow\infty \\
\chi(Y,-1)&=\chi(Y,1)\,\,\,\mathrm{where}\,\,\,\chi=\widehat{V},\widehat{W},{P}_{\mathrm{p}},\frac{\partial \widehat{V}}{\partial Z},\frac{\partial \widehat{W}}{\partial Z}.
\end{align}
This problem has been solved
by \citet{Philip-72} by using conformal maps to find the stream function. The resulting velocity profile is
\begin{align}
\widehat{V}&=-\frac{\partial}{\partial Z}\left(Y\mathrm{Im}\left\{ \frac{1}{\pi}\mathrm{cos}^{-1}\left[\frac{\mathrm{cos}\left(\pi\Theta_{\perp}/2\right)}{\mathrm{cos}\left(\pi\delta/2\right)}\right]\right\} \right)+O\left(\epsilon^{2}\mathrm{Re}\right)\\
\widehat{W}&=\left(Y\frac{\partial}{\partial Y}+1\right)\mathrm{Im}\left\{ \frac{1}{\pi}\mathrm{cos}^{-1}\left[\frac{\mathrm{cos}\left(\pi\Theta_{\perp}/2\right)}{\mathrm{cos}\left(\pi \delta /2 \right)}\right]\right\}+O\left(\epsilon^{2}\mathrm{Re}\right),
\end{align}
where $\Theta_{\perp}=Z+\mathrm{i}Y$. Moreover, it follows from the far-field behavior documented by \citet{Philip-72} for his result that
\begin{equation}
\widehat{V}=O\left(\epsilon^{2}\mathrm{Re}\right),~\widehat{W}=Y+\frac{\lambda}{2}
+O\left(\epsilon^{2}\mathrm{Re}\right)~\mathrm{as}~Y\rightarrow\infty
\end{equation}
such that
\begin{equation}
G\left(\epsilon\right)=\epsilon \lambda +O\left(\epsilon^3\mathrm{Re}\right). \label{eq:geps}
\end{equation}
The complete inner velocity profile becomes
\begin{align}
V&=-2\epsilon\frac{\partial}{\partial Z}\left(Y\mathrm{Im}\left\{ \frac{1}{\pi}\mathrm{cos}^{-1}\left[\frac{\mathrm{cos}\left(\pi\Theta_{\perp}/2\right)}{\mathrm{cos}\left(\pi\delta/2\right)}\right]\right\} \right)+O\left(\epsilon^{3}\mathrm{Re}\right)\label{eq:Trans_inner_V}\\
W&=-\epsilon^{2}Y^{2}+2\epsilon\left(Y\frac{\partial}{\partial Y}+1\right)\mathrm{Im}\left\{ \frac{1}{\pi}\mathrm{cos}^{-1}\left[\frac{\mathrm{cos}\left(\pi\Theta_{\perp}/2\right)}{\mathrm{cos}\left(\pi\delta/2\right)}\right]\right\} +O\left(\epsilon^{3}\mathrm{Re}\right)\label{eq:trans_inner_W}.
\end{align}
The periodic component of the pressure field is not required to resolve the thermal problem and thus not developed here.

\subsubsection{Composite Solution}
As in the case of parallel ridges, the outer solution keeps its form in the overlap region. Therefore, the inner solution is the composite one as per
\begin{align}
\tilde{v}_{\mathrm{comp}} &=-2\epsilon\frac{\partial}{\partial \tilde{z}}\left[\tilde{y}\mathrm{Im}\left\{ \frac{1}{\pi}\mathrm{cos}^{-1}\left[\frac{\mathrm{cos}\left(\pi \left(\tilde{z}/\epsilon + \mathrm{i}\tilde{y}/\epsilon\right)/2\right)}{\mathrm{cos}\left(\pi\delta/2\right)}\right]\right\} \right]+O\left(\epsilon^{3}\mathrm{Re}\right) \\
\tilde{w}_{\mathrm{comp}}&=-\tilde{y}^{2}+2\epsilon\left(\tilde{y}\frac{\partial}{\partial \tilde{y}}+1\right)\mathrm{Im}\left( \frac{1}{\pi}\mathrm{cos}^{-1}\left\{\frac{\mathrm{cos}\left[\pi \left(\tilde{z}/\epsilon + \mathrm{i}\tilde{y}/\epsilon\right)/2\right]}{\mathrm{cos}\left(\pi\delta/2\right)}\right\}\right) +O\left(\epsilon^{3}\mathrm{Re}\right).
\end{align}

\subsubsection{Slip Length}
The procedure for finding the slip length by equating $\tilde{w}_{1\mathrm{d}}$ as per  (\ref{eq:b1d}) and the outer velocity profile in the case of parallel ridges also holds for transverse ones; consequently, 
\begin{equation}
\tilde{b} = \epsilon \lambda/2 + O\left(\epsilon^3\mathrm{Re}\right).
\label{eq:btran}
\end{equation}
This is half of its value for parallel ridges, as is well known, and the error is $O\left(\epsilon^3\mathrm{Re}\right)$ rather
than exponentially small.

\subsubsection{Discussion}
\citet{Teo-09} obtained the same slip length for transverse ridges by taking the limit of the dual-series equations accommodating the mixed boundary condition along the composite interface as $\epsilon \rightarrow 0$, but did not show that the neglect of inertial effects introduces an error term of $O\left(\epsilon^3\mathrm{Re}\right)$. \citet{Davies-06} numerically resolved 
the full problem, including the effect of viscous shear by the gas phase on the liquid. Then, the dimensionless slip length ($\tilde{b}$), in addition to depending on solid fraction ($\phi$) and our small parameter ($\epsilon$), further depends on the nondimensional depth of the cavity, Reynolds number, etc.\footnote{$\epsilon$ need not be small in the numerical study by \citet{Davies-06}.} Such dependencies were explored by \citet{Davies-06} and their results compared reasonably well against experimental data.

\subsection{Thermal Problem}
\subsubsection{Formulation}
Again following \citet{Patankar-77}, the temperature field is decomposed into
\begin{equation}
T=\gamma{z}+T_{\mathrm{p}}\left({y},{z}\right),
\end{equation}
where $\gamma z$ is the linear temperature gradient in the liquid when a uniform heat flux of $\phi q''_{\mathrm{sl}}$ is applied over the composite interface and $T_{\mathrm{p}}\left({y},{z}\right)$ is its periodic component, which repeats itself over
 a period of $2d$. It follows from an energy balance that 
\begin{equation}
\gamma=\frac{q''_{\mathrm{sl}}\phi}{\rho Q' c_{\mathrm{p}}},
\end{equation}
where $Q'$ is the volumetric flow rate of liquid per unit depth of the domain. The dimensional form of the thermal energy equation is
\begin{equation}
v\frac{\partial T_{\mathrm{p}}}{\partial y}+w\left(\gamma+\frac{\partial T_{\mathrm{p}}}{\partial z}\right)=\alpha\left(
\frac{\partial^2 T_{\mathrm{p}}}{\partial y^2}+\frac{\partial ^2T_{\mathrm{p}}}{\partial z^2}\right).
\end{equation}
Unlike in the case of parallel ridges, the axial conduction term is finite. Moreover, we must 
retain it because,
as shown below, it is essential in the inner region. Too, rather than being constant in the domain, the (bulk) mean temperature varies along its streamwise direction as per
\begin{equation}
T_{\mathrm{m}}\left(z\right)=\frac{\int_0^H 
wT\mathrm{d}y}{\int_0^Hw\mathrm{d}y}.
\end{equation}
Consequently, nondimensional temperature is defined relative to the (bulk) mean temperature at the domain inlet as per
\begin{equation}
\tilde{T}=\frac{k\left[T-T_{\mathrm{m}}\left({z}=0\right)\right]}{q_{\mathrm{sl}}''H}.
\end{equation}
We, again, work in terms of $Z$ such that our domain boundaries are independent of $\epsilon$. The dimensionless form of the thermal energy equation governing the periodic component of the temperature field is
\begin{equation}
\mathrm{Pe}\left(\tilde{v}\frac{\partial \tilde{T}_{\mathrm{p}}}{\partial \tilde{y}}+
\frac{\tilde{w}}{\epsilon}\frac{\partial \tilde{T}_{\mathrm{p}}}{\partial Z}\right) +\frac{\tilde{w}\phi}{\tilde{Q}'}=
\frac{\partial^2 \tilde{T}_{\mathrm{p}}}{\partial \tilde{y}^2}+\frac{1}{\epsilon^2}
\frac{\partial^2 \tilde{T}_{\mathrm{p}}}{\partial Z^2},
\label{eq:tebt}
\end{equation}
where the P{\'e}clet number Pe equals RePr, where the Prandtl number $\mathrm{Pr}$ equals $c_{\mathrm{p}}\mu/k$, i.e.,
\begin{equation}
\mathrm{Pe} = \frac{-\rho \beta H^3 c_{\mathrm{p}}}{2\mu k}, 
\end{equation}
and $\tilde{Q}' = 2\mu Q'/\left(-\beta H^3\right)$.
It's subject to the discontinuous (Neumann) boundary condition along the composite interface, symmetry along the channel centerline and periodicity on the upstream and downstream faces of the domain as per 
\begin{align}
\frac{\partial \tilde{T}_{\mathrm{p}}}{\partial{\tilde{y}}}&=0\,\,\,\mathrm{at}\,\,\,{\tilde{y}}=0\,\,\,\mathrm{for}\,\,\,\vert Z\vert<{\delta} \\
\frac{\partial \tilde{T}_{\mathrm{p}}}{\partial{\tilde{y}}}&=-1\,\,\,\mathrm{at}\,\,\,{\tilde{y}}=0\,\,\,\mathrm{for}\,\,\,\vert \delta \vert <\vert Z\vert<1 \\
\frac{\partial \tilde{T}_{\mathrm{p}}}{\partial{\tilde{y}}}&=0\,\,\,\mathrm{at}\,\,\,\tilde{y}=1\,\,\,\mathrm{for}\,\,\,\vert Z\vert<1 \label{eq:adbc}\\
\chi(\tilde{y},-1)&=\chi(\tilde{y},1),\,\,\,\mathrm{where}\,\,\,\chi=\tilde{T}_{\mathrm{p}}, 
\frac{\partial \tilde{T}_{\mathrm{p}}}{\partial \tilde{z}}, \label{eq:pbc}
\end{align}
respectively.

\subsubsection{Outer Region}\label{sec:ortr}
Recalling that, in the outer region, $\tilde{v}$ is exponentially small as per  (\ref{eq:voutert}), the
thermal energy equation becomes 
\begin{equation}
\mathrm{Pe}
\frac{\tilde{w}}{\epsilon}\frac{\partial \tilde{T}_{\mathrm{p}}}{\partial Z}+\frac{\tilde{w}\phi}{\tilde{Q}'} =
\frac{\partial^2 \tilde{T}_{\mathrm{p}}}{\partial \tilde{y}^2}+\frac{1}{\epsilon^2}
\frac{\partial^2 \tilde{T}_{\mathrm{p}}}{\partial Z^2} +\mathrm{e.s.t.}
\end{equation}
Defining 
\begin{equation}
\hat{{T}}_{\mathrm{p}} =\tilde{Q}' \tilde{T}_{\mathrm{p}}
\end{equation}
and recalling that the outer (streamwise) velocity profile is $\tilde{w} = -\tilde{y}^2 + 2\tilde{y} + G\left(\epsilon\right) + \mathrm{e.s.t.}$, it becomes
\begin{equation}
\frac{\mathrm{Pe}}{\epsilon}\left[-\tilde{y}^2 + 2\tilde{y} +G\left(\epsilon\right) \right]
\frac{\partial \hat{{T}}_{\mathrm{p}}}{\partial Z}
+\phi  \left[-\tilde{y}^2 + 2\tilde{y} 
+ G\left(\epsilon \right)\right]
=
\frac{\partial^2 \hat{{T}}_{\mathrm{p}}}{\partial \tilde{y}^2}+\frac{1}{\epsilon^2}
\frac{\partial^2 \hat{{T}}_{\mathrm{p}}}{\partial Z^2} + \mathrm{e.s.t.}
\end{equation}
The boundary conditions are given by  (\ref{eq:adbc}) and, for $\epsilon \ll \tilde{y} <1$,
by  (\ref{eq:pbc}), where $\tilde{T}_{\mathrm{p}}$ is replaced by  $\hat{{T}}_{\mathrm{p}}$.

We, as an ansatz, assume that the periodic component of the temperature field is $O\left(1\right)$ and
thus asymptotically expand it as 
\begin{equation}
\hat{{T}}_{\mathrm{p}}
\sim \sum_{n=0}^{\infty}\epsilon^{n} \hat{{T}}_{\mathrm{p},n}+\mathrm{e.s.t.},
\end{equation}
where $\hat{{T}}_{\mathrm{p,}n}=O(1)$ for $n\geq0$. Then, at the various orders of $\epsilon$, the thermal energy 
equation is
\begin{align}
&O\left(\epsilon^{-2}\right):~\frac{\partial^{2}\hat{{T}}_{\mathrm{p},0}}{\partial Z^{2}}=0 \label{eq:tton2}\\
&O\left(\epsilon^{-1}\right):~
\mathrm{Pe}\,\tilde{w}_0\frac{\partial \hat{{T}}_{\mathrm{p},0}}{\partial Z}
=\frac{\partial^{2}\hat{{T}}_{\mathrm{p},1}}{\partial Z^{2}} \label{eq:tton1}\\
&O\left(\epsilon^n\right):~\mathrm{Pe}\sum_{m=0}^{n+1}\tilde{w}_{n+1-m}\frac{\partial \hat{T}_{\mathrm{p},m}}{\partial Z}
+\phi  \tilde{w}_n
= \frac{\partial^{2}\hat{T}_{\mathrm{p},n}}{\partial \tilde{y}^{2}}
+\frac{\partial^{2}\hat{T}_{\mathrm{p},n+2}}{\partial Z^{2}}~\mathrm{for}~n\geq0,
\end{align}
where the $\tilde{w}_n$ are independent of $Z$ as per  (\ref{eq:tov}).
The periodicity boundary condition applied, in succession, to  (\ref{eq:tton2}) and (\ref{eq:tton1}) shows that 
$\hat{T}_{\mathrm{p},n} = \hat{T}_{\mathrm{p},n}\left(\tilde{y}\right)$ for $n =0,1$. Continuing this process for
the thermal energy equation for $O\left(\epsilon^n\right)$, where $n \geq 0$, shows that
\begin{equation}
\frac{\partial^{2}\hat{T}_{\mathrm{p,}n+2}}{\partial Z^{2}} =\frac{\partial^{2}\hat{T}_{\mathrm{p,}n+2}}{\partial Z^{2}}\left(\mathrm{Pe},\phi,\tilde{w}_0,...,\tilde{w}_{n+1},
\hat{T}_{\mathrm{p,}0},...,\hat{T}_{\mathrm{p,}n+1}
\right).
\end{equation}
Moreover, once the thermal energy equation for order $O\left(\epsilon^n\right)$ has been reached, $\hat{T}_{\mathrm{p},i}$ for $i \leq n+1$ 
has been shown to be only a function of $\tilde{y}$. Therefore, the periodicity boundary condition implies that 
$\hat{T}_{\mathrm{p},n}$ is independent of $Z$ (and thus Pe) for all $n$. Making further use of  (\ref{eq:tov}), 
the thermal energy equation becomes
\begin{align}
&O\left(\epsilon^0\right):~
\frac{\mathrm{d}^{2}\hat{T}_{\mathrm{p},0}}{\mathrm{d}\tilde{y}^{2}}=\phi  \left(-\tilde{y}^2 + 2\tilde{y} 
+ G_0\right)\\
&O\left(\epsilon^n\right):~\frac{\mathrm{d}^{2}\hat{T}_{\mathrm{p},n}}{\mathrm{d}\tilde{y}^{2}}=\phi G_n,~n \geq 1.
\end{align}
Integrating these equations, applying the symmetry boundary condition at $\tilde{y} =1 $ and integrating them again yields,
\begin{align}
&O\left(\epsilon^0\right):~\hat{T}_{\mathrm{p},0} =\phi 
\left[ 
-\frac{\left(\tilde{y}-1\right)^4}{12} +\frac{\left(\tilde{y}-1\right)^2}{2} + G_0 \frac{\left(\tilde{y}-1\right)^2}{2} 
\right] + \alpha_0 \\
&O\left(\epsilon^n\right):~\hat{T}_{\mathrm{p},n} = \phi G_n \frac{\left(\tilde{y}-1\right)^2}{2}  + \alpha_n,~n \geq 1,
\end{align}
where the $\alpha_n$ are constants. Consequently,
\begin{equation}
\hat{T}_{\mathrm{p}} = \phi 
\left[ 
-\frac{\left(\tilde{y}-1\right)^4}{12} +\frac{\left(\tilde{y}-1\right)^2}{2} + G\left(\epsilon\right)\frac{\left(\tilde{y}-1\right)^2}{2}
\right]+\sum_{n=0}^{\infty}\alpha_n \epsilon^n +\mathrm{e.s.t.}\label{eq:outertranT}
\end{equation}
Finally, it follows from the slip length given by  (\ref{eq:btran}) that $\tilde{Q}' =  2/3 + \epsilon \lambda + 
O\left(\epsilon^3\mathrm{Re}\right)$ and thus 
\begin{equation}
\frac{1}{\tilde{Q}'} = \frac{1}{2/3 +\epsilon \lambda} + O\left(\epsilon^3\mathrm{Re}\right)
\end{equation}
and, as per  ({\ref{eq:geps}), 
$G\left(\epsilon\right)=\epsilon \lambda +O\left(\epsilon^3\mathrm{Re}\right)$. Therefore, reverting 
from $\hat{T}_{\mathrm{p}}$ back to $\tilde{T}_{\mathrm{p}}$,  (\ref{eq:outertranT}) is expressed as 
\begin{equation}
\tilde{T}_{\mathrm{p}}=\frac{\phi}{2/3+\epsilon \lambda}\left[-\frac{\left(\tilde{y}-1\right)^{4}}{12}+\left(\frac{1}{2}+\frac{\epsilon \lambda}{2}\right)\left(\tilde{y}-1\right)^{2}+H\left(\epsilon\right)+\mathrm{e.s.t.}\right]
+O\left(\epsilon^3\mathrm{Re}\right),
\label{eq:outTt}
\end{equation}
where $H_n = \alpha_n/\phi$ and
\begin{equation}
H\left(\epsilon\right)=\sum_{n=0}^{\infty} H_n \epsilon^n.\label{eq:otptr}
\end{equation}
This outer temperature profile is the same as that for parallel ridges as per  (\ref{eq:poutT}), 
except that the (asymptotic limit of the) slip length for transverse ridges ($\epsilon \lambda/2$) replaces that for parallel ones ($\epsilon \lambda$) and the error is
$O\left(\epsilon^3\mathrm{Re}\right)$ rather than exponentially small.

\subsubsection{Inner Region}\label{sec:itt}
Denoting $\tilde{T}_{\mathrm{p}}$ by ${\theta}_{\mathrm{p}}\left(Y,Z\right)$ in the inner region, 
the thermal energy equation is
\begin{equation}
\frac{\mathrm{Pe}}{\epsilon}\left(V\frac{\partial{\theta}_{\mathrm{p}}}{\partial Y}
+W\frac{\partial{\theta}_{\mathrm{p}}}{\partial Z}\right)+
\frac{W\phi}{\tilde{Q}'}
=\frac{1}{\epsilon^{2}}\left(\frac{\partial^{2}{\theta}_{\mathrm{p}}}{\partial Y^{2}}+\frac{\partial^{2}{\theta}_{\mathrm{p}}}{\partial Z^{2}}\right),\label{eq:trte}
\end{equation}
where $V$ and $W$ are given by  (\ref{eq:Trans_inner_V}) and (\ref{eq:trans_inner_W}), respectively, and 
are $O\left(\epsilon\right)$. 
The boundary conditions are
\begin{align}
\frac{\partial\theta_{\mathrm{p}}}{\partial Y}&=0~\mathrm{at}~Y=0~\mathrm{for}~0<Z<\delta\\
\frac{\partial\theta_{\mathrm{p}}}{\partial Y}&=-\epsilon~\mathrm{at}~Y=0~\mathrm{for}~\delta<Z<1 \label{eq:qppc} \\
\theta_{\mathrm{p}}&\sim-\epsilon\phi Y+\phi\left[\frac{1}{2}+\frac{1/12+H(\epsilon)}{2/3+\epsilon \lambda}\right]
+O\left(\epsilon^{3},\epsilon^3\mathrm{Re}\right)\,\,\mathrm{as}\,\, Y\rightarrow\infty \label{eq:trpm}\\
\chi(\tilde{y},Z)&=\chi(\tilde{y},Z+2),\,\,\,\mathrm{where}\,\,\,\chi=\theta_{\mathrm{p}}, 
\frac{\partial\theta_{\mathrm{p}}}{\partial \tilde{z}}, \label{eq:pbt}
\end{align}
where  (\ref{eq:trpm}), which follows from the manipulation of  (\ref{eq:outTt}), is the matching condition in the overlap region.
As an ansatz, we assume that the  periodic component of the dimensionless, inner temperature field is $O\left(1\right)$ and
thus asymptotically expand it as 
\begin{equation}
{\theta}_{\mathrm{p}}
\sim \sum_{n=0}^{\infty}\epsilon^{n} {\theta}_{\mathrm{p},n}+\mathrm{e.s.t.},
\end{equation}
The thermal energy equation reduces to Laplace's equation at leading-order, i.e., $O\left(\epsilon^{-2}\right)$; therefore, 
\begin{equation}
\frac{\partial \theta_{\mathrm{p},0}}{\partial Y} \sim \frac{\partial \theta_{\mathrm{p},0}}{\partial Z}.
\end{equation}
These quantities do not exceed their values near the ridge and are thus $O\left(\epsilon\right)$ as per the boundary condition along it as 
per  (\ref{eq:qppc}) and $V,W=O\left(\epsilon\right)$; consequently, the thermal energy equation reduces to
\begin{equation}
\frac{\partial^{2}\theta_{\mathrm{p}}}{\partial Y^{2}}+\frac{\partial^{2}\theta_{\mathrm{p}}}{\partial Z^{2}}=
O\left(\epsilon^3\mathrm{Pe},\epsilon^3\right).
\label{eq:oi}
\end{equation}
Accepting an accuracy of $O\left(\epsilon^{2}\right)$ for $\theta_{\mathrm{p}}$, the periodic boundary conditions along the
streamwise borders of the domain manifest themselves as symmetry conditions on a diffusion problem such that 
 (\ref{eq:pbt}) is replaced by 
\begin{equation}
\frac{\partial\theta_{\mathrm{p}}}{\partial Z}=0~\mathrm{at}~\left|Z\right|=1.
\end{equation}
Relatedly, since the outer problem has no dependence on $Z$, the temperature field and thus local Nusselt number are 
symmetric about $Z = 0$ and we henceforth need only consider our domain to extend from 0 to $1$. 
The solution given by \citet{Mikic-57} is again employed such that the dimensionless, inner
periodic temperature profile is given by
\begin{equation}
\theta_{\mathrm{p}} = -\epsilon \phi Y  - \frac{2\epsilon}{\pi^2}\sum_{n=1}^{\infty}\frac{\sin\left(n\pi\delta\right)\cos\left(n\pi Z\right)\mathrm{e}^{-n \pi Y}}{n^2}+\phi\left[\frac{1}{2}+\frac{1/12+H(\epsilon)}{2/3+\epsilon \lambda}\right]+
O\left(\epsilon^3,\epsilon^3\mathrm{Re},\epsilon^3\mathrm{Pe}\right).
\label{eq:mikic2}
\end{equation}

Turning our attention to finding $H\left(\epsilon\right)$, we first observe that, as per a (dimensional) energy balance on a control
volume bounded by $z = 0$ and some arbitrary $z$ in the streamwise direction and spanning the height of the domain, that
\begin{equation}
T_{\mathrm{m}}\left(z\right) - T_{\mathrm{m}}\left(0\right) = \frac{1}{\rho Q'c_{\mathrm{p}}}\left[
\int_0^z q''_{\mathrm{c,l}}\left(z\right) \mathrm{d} z - k\int_0^H \frac{\partial T}{\partial z}\left(y,0\right)\mathrm{d} y
+  k\int_0^H \frac{\partial T}{\partial z}\left(y,z\right) \mathrm{d} y
\right],
\end{equation}
where $q''_{\mathrm{c,l}}\left(z\right)$ is the (local) heat flux along the composite interface, i.e., $q''_{\mathrm{sl}}$ along the ridge and 0 along the meniscus. The linear components of $T_{\mathrm{m}}$ on the left side of this equation sum to $\gamma z$ and those in the axial diffusion terms on the right one cancel; consequently,
\begin{align}
T_{\mathrm{m,p}}\left(z\right) - T_{\mathrm{m,p}}\left(0\right) =& \frac{1}{\rho Q'c_{\mathrm{p}}}\left[
\int_0^z q''_{\mathrm{c,l}}\left(z\right) \mathrm{d} z - 
k\int_0^H \frac{\partial T_{\mathrm{p}}}
{\partial z}\left(y,0\right)\mathrm{d} y
+  k\int_0^H \frac{\partial T_{\mathrm{p}}}{\partial z}\left(y,z\right) \mathrm{d} y \right. \nonumber \\  &\left. - q''_{\mathrm{sl}}\phi z
\right],
\end{align}
or, noting that $T_{\mathrm{m}}(0) = T_{\mathrm{m,p}}(0)$ and the axial diffusion terms are e.s.t. in the outer region, in terms of dimensionless inner variables, 
\begin{align}
\tilde{T}_{\mathrm{m,p}}\left({Z}\right)  =& \frac{1}{\tilde{Q}'\mathrm{Pe}}\left[\epsilon
\int_0^{Z} \hat{q}''_{\mathrm{c,l}}\left(Z\right) \mathrm{d}Z - \phi \epsilon Z
- \int_0^{\frac{\eta}{\epsilon}} \frac{\partial\theta_{\mathrm{p}}}{\partial Z}\left(Y,0\right)\mathrm{d} Y
+ \int_0^{\frac{\eta}{\epsilon}} \frac{\partial\theta_{\mathrm{p}}}{\partial Z}\left(Y,Z\right)\mathrm{d} Y
\right] \nonumber \\
 & +\mathrm{e.s.t.},
\label{eq:bulkTt}
\end{align}
where $\hat{q}''_{\mathrm{c,l}}\left(\tilde{z}\right)=q''_{\mathrm{c,l}}/q_{\mathrm{sl}}''$, i.e., 0 along the meniscus and 1 along the ridge, and, as above, $\eta$ is a 
value of $\tilde{y}$ in the overlap region. We note that the $O\left(\epsilon^3\mathrm{Re}\right)$ error term in the (periodic component of the) outer temperature profile (applicable in the overlap region) given by  (\ref{eq:outTt}) relates to that in the volumetric flow rate per unit 
depth through the domain and thus is
independent of $Z$. Hence, the axial-conduction terms need not be integrated beyond an upper limit $\eta/\epsilon$ of in  \eqref{eq:bulkTt}. 

To simplify  (\ref{eq:bulkTt}), we consider a thermal energy balance on a control volume bounded by 
${Z} = 0$ and some arbitrary ${Z}$ in the streamwise direction and by the base of the domain and a value of $\tilde{y}$ in 
the overlap
region. Noting the error introduced by assuming pure diffusion as per  \eqref{eq:oi} and integrating over the control
volume yields 
\begin{align}
- \int_0^{\frac{\eta}{\epsilon}} \frac{\partial\theta_{\mathrm{p}}}{\partial Z}\left(Y,0\right)\mathrm{d} Y
+ \int_0^{\frac{\eta}{\epsilon}} \frac{\partial\theta_{\mathrm{p}}}{\partial Z}\left(Y,Z\right)\mathrm{d} Y+
\epsilon
\int_0^{Z} \hat{q}''_{\mathrm{c,l}}\left(Z\right) \mathrm{d}Z & \nonumber \\
+\int_0^Z\frac{\partial \theta_{\mathrm{p}}}{\partial Y}\left(\frac{\eta}{\epsilon},Z\right) \mathrm{d}Y &=
O\left(\frac{\eta}{\epsilon}\epsilon^3\mathrm{Pe},\frac{\eta}{\epsilon}\epsilon^3\right)
\label{eq:ebY}
\end{align}
Moreover, manipulation of  \eqref{eq:outTt} and, subsequently, integration yields
\begin{equation}\int_0^Z
\frac{\partial \theta_{\mathrm{p}}}{\partial Y}\left(\frac{\eta}{\epsilon},Z\right)\mathrm{d}Z=-\epsilon \phi Z
+O\left(\epsilon^3\frac{\eta}{\epsilon},\epsilon^3\mathrm{Re}\right)
\end{equation}
Subsituting this result into  \eqref{eq:ebY}, it follows that  \eqref{eq:bulkTt} becomes
\begin{equation}
\tilde{T}_{\mathrm{m,p}}\left({Z}\right)  =\frac{1}{\tilde{Q}'\mathrm{Pe}}\left[
O\left(\frac{\eta}{\epsilon}\epsilon^3\mathrm{Pe},\frac{\eta}{\epsilon}\epsilon^3,\epsilon^3\mathrm{Re}\right)
\right]
\end{equation}
This reesult is independent of $\eta$, where $\epsilon \ll \eta \ll 1$; therefore, minimizing it,
\begin{equation}
\tilde{T}_{\mathrm{m,p}}\left({Z}\right)  = O\left(\epsilon^3,\frac{\epsilon^3}{\mathrm{Pe}},\frac{\epsilon^3}{\mathrm{Pr}}\right).
\end{equation}
i.e.,
\begin{equation}
\int_0^1 \tilde{w}\tilde{T}_{\mathrm{p}}\mathrm{d}\tilde{y} =O\left(\epsilon^3,\frac{\epsilon^3}{\mathrm{Pe}},\frac{\epsilon^3}{\mathrm{Pr}}\right).
\end{equation}
Integrating across the domain yields
\begin{equation}
\int_0^1\int_0^{\epsilon} \tilde{w}\tilde{T}_{\mathrm{p}}\mathrm{d}\tilde{z}\mathrm{d}\tilde{y} = 
O\left(\epsilon^4,\frac{\epsilon^4}{\mathrm{Pe}},\frac{\epsilon^4}{\mathrm{Pr}}\right).
\end{equation}
or
\begin{equation}
\int_0^{\xi}\int_0^{\epsilon} \tilde{w}_{\mathrm{in}}\tilde{T}_{\mathrm{p,in}}\mathrm{d}\tilde{z}\mathrm{d}\tilde{y}
+\int_0^1\int_0^{\epsilon} \tilde{w}_{\mathrm{out}}\tilde{T}_{\mathrm{p,out}}\mathrm{d}\tilde{z}\mathrm{d}\tilde{y} 
-\int_0^{\xi}\int_0^{\epsilon} \tilde{w}_{\mathrm{out}}\tilde{T}_{\mathrm{p,out}}\mathrm{d}\tilde{z}\mathrm{d}\tilde{y} = 
O\left(\epsilon^4,\frac{\epsilon^4}{\mathrm{Pe}},\frac{\epsilon^4}{\mathrm{Pr}}\right),
 \label{eq:ebT}
\end{equation}
where $\xi$ is a value of $\tilde{y}$ in the overlap region. 

The first term in the preceding equation may be written as
\begin{align}
\int_0^{\xi}\int_0^{\epsilon} \tilde{w}_{\mathrm{in}}\tilde{T}_{\mathrm{p,in}}\mathrm{d}\tilde{z}\mathrm{d}\tilde{y} &=
\int_0^{\xi}\int_0^{\epsilon} \left[
-\tilde{y}^{2}+\epsilon\left(\tilde{y}\frac{\partial}{\partial \tilde{y}}+1\right)\mathrm{Im}\left(\frac{2}{\pi}\mathrm{cos}^{-1}
\left\{\frac{\mathrm{cos}\left[\frac{\pi}{2} \left(\frac{\tilde{z}}{\epsilon} + \mathrm{i}\frac{\tilde{y}}{\epsilon}\right)\right]}{\mathrm{cos}\left(\frac{\pi\delta}{2}\right)}\right\}
\right) \right. \nonumber \\
&\left.+O\left(\epsilon^{3}\mathrm{Re}\right)
\right] \left\{
 -\phi \tilde{y}  - \frac{2\epsilon}{\pi^2}\sum_{n=1}^{\infty}\frac{\sin\left(n\pi\delta\right)\cos\left(n\pi 
 \tilde{z}/\epsilon\right)\mathrm{e}^{-n \pi 
 \tilde{y}/\epsilon}}{n^2}\right.\nonumber \\
 &\left. + \phi\left[\frac{1}{2}+\frac{1/12+H(\epsilon)}{2/3+\epsilon \lambda}\right]+
O\left(\epsilon^3,\epsilon^3\mathrm{Re},\epsilon^3\mathrm{Pe}\right)\right\}
\mathrm{d}\tilde{z}\mathrm{d}\tilde{y},
\end{align}
Expanding products in the integrand on the right side and quantifying the error for some of the terms upon integration,
\begin{align}
\int_0^{\xi}\int_0^{\epsilon} \tilde{w}_{\mathrm{in}}\tilde{T}_{\mathrm{p,in}}\mathrm{d}\tilde{z}\mathrm{d}\tilde{y} &=
\int_0^{\xi}\int_0^{\epsilon}
\left\{ -\phi \tilde{y}+ \phi\left[\frac{1}{2}+\frac{1/12+H(\epsilon)}{2/3+\epsilon \lambda}\right]
+
O\left(\epsilon^3,\epsilon^3\mathrm{Re},\epsilon^3\mathrm{Pe}\right)\right\}\nonumber \\
&
\left[-\tilde{y}^2+\epsilon\left(\tilde{y}\frac{\partial}{\partial \tilde{y}}+1\right)\mathrm{Im}\left(\frac{2}{\pi}\mathrm{cos}^{-1}
\left\{\frac{\mathrm{cos}\left[\frac{\pi}{2} \left(\frac{\tilde{z}}{\epsilon} + \mathrm{i}\frac{\tilde{y}}{\epsilon}\right)\right]}{\mathrm{cos}\left(\frac{\pi\delta}{2}\right)}\right\}
\right)\right]\mathrm{d}\tilde{z}\mathrm{d}\tilde{y}\nonumber \\
&
+ O\left(\epsilon^4\xi^2\mathrm{Re}\right)
+O\left(\epsilon^5\xi\mathrm{Re}\right) + O\left(\epsilon^4\xi\mathrm{Re}\right)
+ 
\nonumber \\
&O\left(\epsilon^7 \xi \mathrm{Re},\epsilon^7\xi\mathrm{Re}^2,\epsilon^7\xi\mathrm{Re}\,\mathrm{Pe}\right)+O\left(\epsilon^2\xi^3\right)+O\left(\epsilon^3\xi\right)\label{eq:inter}
\end{align}
Not yet making any assumptions about $\xi$, other than it is less than unity, $\mathrm{Re}$ and 
$\mathrm{Pe}$, we proceed by removing several error terms that are small relative to those we keep. Subsequently,
distributing the 
$O\left(\epsilon^3,\epsilon^3\mathrm{Re},\epsilon^3\mathrm{Pe}\right)$ term, 
 \eqref{eq:inter} may be written as
 \begin{align}
\int_0^{\xi} \int_0^{\epsilon}\tilde{w}_{\mathrm{in}}\tilde{T}_{\mathrm{p,in}}\mathrm{d}\tilde{z}\mathrm{d}\tilde{y} &=
\int_0^{\xi}
\left\{ -\phi \tilde{y}+ \phi\left[\frac{1}{2}+\frac{1/12+H(\epsilon)}{2/3+\epsilon \lambda}\right]
\right\}\nonumber \\
&
\left[-\tilde{y}^2\int_0^{\epsilon}\mathrm{d}\tilde{z}+\epsilon\left(\tilde{y}\frac{\partial}{\partial \tilde{y}}+1\right)\int_0^{\epsilon}\mathrm{Im}\left(\frac{2}{\pi}\mathrm{cos}^{-1}
\left\{\frac{\mathrm{cos}\left[\frac{\pi}{2} \left(\frac{\tilde{z}}{\epsilon} + \mathrm{i}\frac{\tilde{y}}{\epsilon}\right)\right]}{\mathrm{cos}\left(\frac{\pi\delta}{2}\right)}\right\}
\right)\mathrm{d}\tilde{z}\right]\mathrm{d}\tilde{y}\nonumber \\
&+O\left(\epsilon^4\xi^2\mathrm{Re}\right)+O\left(\epsilon^4\xi\mathrm{Re}\right)
+ O\left(\epsilon^7\xi\mathrm{Re}^2,\epsilon^7\xi\mathrm{Re}\,\mathrm{Pe}\right)+O\left(\epsilon^2\xi^3\right)
 \nonumber \\
&+O\left(\epsilon^3\xi\right) + O\left(\epsilon^4\xi^3,\epsilon^4\xi^3\mathrm{Re},\epsilon^4\xi^3\mathrm{Pe}\right)+
O\left(\epsilon^5\xi,\epsilon^5\xi\mathrm{Re},\epsilon^5\xi\mathrm{Pe}\right)
\end{align}
The imaginary function in the preceding equation equals  $\widehat{W}$ in the parallel
ridge problem as per  (\ref{eq:What}) when $\tilde{x}$ is replaced by $\tilde{z}$. Consequently, when it's integrated 
across the width of the domain, it becomes $\tilde{y} + \epsilon \lambda$. After deleting relatively small error terms, it follows that
\begin{align}
\int_0^{\xi} \int_0^{\epsilon}\tilde{w}_{\mathrm{in}}\tilde{T}_{\mathrm{p,in}}\mathrm{d}\tilde{z}\mathrm{d}\tilde{y} &=
\int_0^{\xi}\int_0^{\epsilon}\left(-\tilde{y}^2+2\tilde{y}+\epsilon \lambda\right)
\left\{ -\phi \tilde{y}+ \phi\left[\frac{1}{2}+\frac{1/12+H(\epsilon)}{2/3+\epsilon \lambda}\right]
\right\}
\mathrm{d}\tilde{z}\mathrm{d}\tilde{y}\nonumber \\
&+O\left(\epsilon^4\xi^2\mathrm{Re}\right)+O\left(\epsilon^4\xi\mathrm{Re}\right)+ O\left(\epsilon^7\xi\mathrm{Re}^2,\epsilon^7\xi\mathrm{Re}\,\mathrm{Pe}\right)+O\left(\epsilon^2\xi^3\right)\nonumber \\
&
+O\left(\epsilon^3\xi\right)
+O\left(\epsilon^4\xi^3,\epsilon^4\xi^3\mathrm{Re},\epsilon^4\xi^3\mathrm{Pe}\right)+
O\left(\epsilon^5\xi\mathrm{Pe}\right)
\end{align}
Setting $\xi = \epsilon$,
\begin{align}
\int_0^{\xi} \int_0^{\epsilon}\tilde{w}_{\mathrm{in}}\tilde{T}_{\mathrm{p,in}}\mathrm{d}\tilde{z}\mathrm{d}\tilde{y} &=
\int_0^{\xi}\int_0^{\epsilon}\left(-\tilde{y}^2+2\tilde{y}+\epsilon \lambda\right)
\left\{ -\phi \tilde{y}+ \phi\left[\frac{1}{2}+\frac{1/12+H(\epsilon)}{2/3+\epsilon \lambda}\right]
\right\}
\mathrm{d}\tilde{z}\mathrm{d}\tilde{y}\nonumber \\
&O\left(\epsilon^5\mathrm{Re}\right) 
+ O\left(\epsilon^8\mathrm{Re}^2,\epsilon^8\mathrm{Re}\,\mathrm{Pe}\right)+O\left(\epsilon^4\right)
+O\left(\epsilon^6\mathrm{Pe}\right)
\label{eq:infT}
\end{align}

Turning our attention to the (negative of the) third term on the left side of  (\ref{eq:ebT}), it follows from 
 (\ref{eq:tov}), (\ref{eq:geps}) and (\ref{eq:trpm}) that
\begin{align}
\int_0^{\xi}\int_0^{\epsilon} \tilde{w}_{\mathrm{out}}\tilde{T}_{\mathrm{p,out}}\mathrm{d}\tilde{z}\mathrm{d}\tilde{y} 
&=
\int_0^{\xi}\int_0^{\epsilon}\left(-\tilde{y}^2+2\tilde{y}+\epsilon \lambda\right)
\left\{ -\phi \tilde{y}+ \phi\left[\frac{1}{2}+\frac{1/12+H(\epsilon)}{2/3+\epsilon \lambda}\right]
\right\}
\mathrm{d}\tilde{z}\mathrm{d}\tilde{y}\nonumber \\
&
+O\left(\epsilon^4\xi^2\mathrm{Re}\right)+O\left(\epsilon^4\xi\mathrm{Re}\right)+O\left(\epsilon^7\xi\mathrm{Re},\epsilon^7\xi\mathrm{Re}^2\right)\nonumber \\
&+O\left(\epsilon^4\xi^3,\epsilon^4\xi^3\mathrm{Re}\right)
+O\left(\epsilon^4\xi^2,\epsilon^4\xi^2\mathrm{Re}\right)+O\left(\epsilon^5\xi,\epsilon^5\xi\mathrm{Re}\right)
\end{align}
Removing error terms that are small relative to those we keep and setting $\xi$ equal to $\epsilon$,
\begin{align}
\int_0^{\xi}\int_0^{\epsilon} \tilde{w}_{\mathrm{out}}\tilde{T}_{\mathrm{p,out}}\mathrm{d}\tilde{z}\mathrm{d}\tilde{y} 
&=
\int_0^{\xi}\int_0^{\epsilon}\left(-\tilde{y}^2+2\tilde{y}+\epsilon \lambda\right)
\left\{ -\phi \tilde{y}+ \phi\left[\frac{1}{2}+\frac{1/12+H(\epsilon)}{2/3+\epsilon \lambda}\right]
\right\}
\mathrm{d}\tilde{z}\mathrm{d}\tilde{y}\nonumber \\
&O\left(\epsilon^6\right)
+O\left(\epsilon^5\mathrm{Re}\right)+O\left(\epsilon^8\mathrm{Re}^2\right).
\label{eq:outinT}
\end{align}

Substituting the results given by  (\ref{eq:infT}) and~(\ref{eq:outinT}) into  (\ref{eq:ebT}), it becomes
\begin{equation}
\int_0^1\int_0^{\epsilon}  \tilde{w}_{\mathrm{out}}\tilde{T}_{\mathrm{p,out}}\mathrm{d}\tilde{y}
\mathrm{d}\tilde{z}=
O\left(
\frac{\epsilon^4}{\mathrm{Pe}},
 \frac{\epsilon^4}{\mathrm{Pr}},\epsilon^4,\epsilon^5\mathrm{Re},\epsilon^6\mathrm{Pe},\epsilon^8\mathrm{Re}^2,
\epsilon^8\mathrm{Re}\,\mathrm{Pe}
 \right).
\end{equation}
Moreover, it follows from our results for $\tilde{w}$ and $\tilde{T}$ that
\begin{align}
\int_0^1\int_0^{\epsilon} \tilde{w}_{\mathrm{out}}\tilde{T}_{\mathrm{p,out}}\mathrm{d}\tilde{z}\mathrm{d}\tilde{y}&=
\int_0^1\int_0^{\epsilon}\left(-\tilde{y}^2+2\tilde{y}+\epsilon \lambda\right)\frac{\phi}{2/3+\epsilon \lambda}
\left[-\frac{\left(\tilde{y}-1\right)^{4}}{12}\right. \nonumber \\ &\left.+\left(\frac{1}{2}+\frac{\epsilon \lambda}{2}\right)\left(\tilde{y}-1\right)^{2}+H\left(\epsilon\right)\right]\mathrm{d}\tilde{z}\mathrm{d}\tilde{y}+O\left(\epsilon^4\mathrm{Re}\right).
\end{align}

This result may be expressed as
\begin{align}
&\int_0^1 \left(-\tilde{y}^2+2\tilde{y}+\epsilon \lambda\right)
\frac{\phi}{2/3+\epsilon \lambda}\left[-\frac{\left(\tilde{y}-1\right)^{4}}{12}+\left(\frac{1}{2}+\frac{\epsilon \lambda}{2}\right)\left(\tilde{y}-1\right)^{2}+H\left(\epsilon\right)\right]\mathrm{d}\tilde{y} = \nonumber \\
&O\left(
\frac{\epsilon^3}{\mathrm{Pe}},
 \frac{\epsilon^3}{\mathrm{Pr}},\epsilon^3,\epsilon^4\mathrm{Re},\epsilon^5\mathrm{Pe},
\epsilon^7\mathrm{Re}\,\mathrm{Pe}
 \right).
\end{align}

Referring back to the corresponding result for parallel ridges, i.e.,  (\ref{eq:bulkint}), it's apparent that
\begin{equation}
H\left(\epsilon \right)=\breve{H}+O\left(
\frac{\epsilon^3}{\mathrm{Pe}},
 \frac{\epsilon^3}{\mathrm{Pr}},\epsilon^3,\epsilon^4\mathrm{Re},\epsilon^5\mathrm{Pe},
\epsilon^7\mathrm{Re}\,\mathrm{Pe},
 \right)
\end{equation}
where
\begin{equation}
\breve{H}\left(\epsilon \right)= -\frac{1}{140}\frac{13+91\epsilon \lambda/2 + 35 \left(\epsilon \lambda\right)^2}{1+3\epsilon \lambda/2}
\label{eq:heps}.
\end{equation}
i.e., it's the same as $D\left(\epsilon\right)$, except that the leading order slip length for parallel ridges $\left(\epsilon \lambda\right)$ 
is replaced by that  for transverse ones $\left(\epsilon \lambda/2\right)$ and inertial and transverse advection effects may increase the error. By further analogy with the parallel ridge problem, we define 
\begin{equation}
\frac{1}{2}+\frac{1/12+H(\epsilon)}{2/3+\epsilon \lambda}= \hat{H}\left(\epsilon \lambda\right)+
O\left(
\frac{\epsilon^3}{\mathrm{Pe}},
 \frac{\epsilon^3}{\mathrm{Pr}},\epsilon^3,\epsilon^4\mathrm{Re},\epsilon^5\mathrm{Pe},
\epsilon^7\mathrm{Re}\,\mathrm{Pe}
 \right)
\label{eq:hstardef}
\end{equation}
where
\begin{equation}
\hat{H}\left(\epsilon \lambda\right) = \frac{17+42\epsilon \lambda + 105 \left(\epsilon \lambda\right)^2/4}
{35\left(1+3\epsilon \lambda /2\right)^2},
\label{eq:hhat}
\end{equation}
i.e., $\hat{H}\left(\epsilon \lambda\right)  =\hat{D}\left(\epsilon \lambda/2\right)$ from the parallel 
ridge problem and too varies between 1/3 and 17/35. The temperature along the composite interface becomes
\begin{equation}
\theta \left(X,0\right) = \phi \hat{H}\left(\epsilon \lambda\right)  - \frac{2\epsilon}{\pi^2}\sum_{n=1}^{\infty}\frac{\sin\left(n\pi\delta\right)\cos\left(n\pi Z\right)}{n^2}+
O\left(
\frac{\epsilon^3}{\mathrm{Pe}},
 \frac{\epsilon^3}{\mathrm{Pr}},\epsilon^3,\epsilon^3\mathrm{Re},\epsilon^5\mathrm{Pe},
\epsilon^7\mathrm{Re}\,\mathrm{Pe}
 \right).
\label{eq:thetacpT}
\end{equation}

\subsubsection{Composite Solution}
The composite solution follows from  (\ref{eq:tcomp}) and includes the linear component of the temperature field. It is
\begin{align}
\tilde{T}_{\mathrm{comp}} &= 
\frac{\phi}{2/3+\epsilon \lambda}\left[\frac{\tilde{z}}{\mathrm{Pe}}
-\frac{\left(\tilde{y}-1\right)^{4}}{12}+\left(\frac{1}{2}+\frac{\epsilon \lambda}{2}\right)\left(\tilde{y}-1\right)^{2}+\breve{H}\left(\epsilon\right)\right]  -
\nonumber \\
& \frac{2\epsilon}{\pi^2}\sum_{n=1}^{\infty}\frac{\sin\left(n\pi\delta\right)\cos\left(n\pi \tilde{z}/\epsilon\right)
\mathrm{e}^{-n \pi \tilde{y}/\epsilon}}{n^2}
+O\left(
\frac{\epsilon^3}{\mathrm{Pe}},
 \frac{\epsilon^3}{\mathrm{Pr}},\epsilon^3,\epsilon^4\mathrm{Re},\epsilon^3\mathrm{Pe},
\epsilon^7\mathrm{Re}\,\mathrm{Pe}
 \right).
\end{align}

\subsubsection{Nusselt Numbers}
The Nusselt number along the ridge follows from  (\ref{eq:nudef}) and (\ref{eq:thetacpT}) as
\begin{equation}
\mathrm{Nu} =
\frac{4}{\phi \hat{D}\left(\epsilon \lambda/2\right) - \frac{2\epsilon}{\pi^2}
\sum_{n=1}^{\infty}\frac{\sin\left(n\pi\delta\right)\cos\left(n\pi Z\right)}{n^2}}
+O\left(\frac{\epsilon^3}{\mathrm{Pe}},
 \frac{\epsilon^3}{\mathrm{Pr}},\epsilon^3,\epsilon^4\mathrm{Re},\epsilon^5\mathrm{Pe},
\epsilon^7\mathrm{Re}\,\mathrm{Pe} \right)
\label{eq:nuloct}
\end{equation}
and, as in the case of parallel ridges, approaches $\infty$ as solid fraction approaches 0. Regularization
of this expression and,
subsequently, an expansion of it in terms of $\epsilon/[\hat{D}\left(\epsilon \lambda/2\right)+\epsilon \lambda]$ follows in the same manner as in the case of parallel ridges. The result is that the mean Nusselt number $\overline{\mathrm{Nu}}$ is given by  (\ref{eq:nupf}) when 
$\hat{D} \left(\epsilon \lambda\right)$ is replaced by $\hat{D}\left(\epsilon \lambda/2\right)$ and the error term is replaced by that 
in  (\ref{eq:nuloct}). Insofar as the alternate definition of the mean Nusselt number, it follows that, by replacing 
$\epsilon \lambda$ with $\epsilon \lambda/2$ in  (\ref{eq:nupp}) and adjusting the error term that
\begin{align}
\overline{\mathrm{Nu}}'= &140\left(1+\frac{3\epsilon \lambda}{2}\right)^2/\left\{
17+ \left[42\lambda + \frac{70}{\phi^2\pi^3}\sum_{n=1}^{\infty}\frac{\sin^2\left(n\pi\delta\right)}{n^3}\right]\epsilon \right. \nonumber \\
&\left.+\left[
\frac{105\lambda^2}{4} +\frac{210\lambda}{\phi^2\pi^3}\sum_{n=1}^{\infty}\frac{\sin^2\left(n\pi\delta\right)}{n^3}
\right]\epsilon^2
 +\left[ \frac{630\lambda^2}{4\phi^2\pi^3}\sum_{n=1}^{\infty}\frac{\sin^2\left(n\pi\delta\right)}{n^3}\right]\epsilon^3\right\} \nonumber \\
 &+
 O\left(
\frac{\epsilon^3}{\mathrm{Pe}},
 \frac{\epsilon^3}{\mathrm{Pr}},\epsilon^3,\epsilon^4\mathrm{Re},\epsilon^5\mathrm{Pe},
\epsilon^7\mathrm{Re}\,\mathrm{Pe}
 \right).
\end{align}
In summary, Table~\ref{tab:ps} applies transverse ridges when $\hat{D} \left(\epsilon \lambda\right)$ is replaced by $\hat{D}\left(\epsilon \lambda/2\right)$ in the Nusselt number expressions and the $O(\epsilon^3)$ error terms are replaced by $O\left(
\frac{\epsilon^3}{\mathrm{Pe}},
 \frac{\epsilon^3}{\mathrm{Pr}},\epsilon^3,\epsilon^4\mathrm{Re},\epsilon^5\mathrm{Pe},
\epsilon^7\mathrm{Re}\,\mathrm{Pe}
 \right)$.

\subsubsection{Results}
We plot $\overline{\mathrm{Nu}}$ versus $\phi$ for parallel ridges (in green based on our 
asymptotic result,
 \eqref{eq:nupp} and in black based on the exact result by \citet{Kirk-17}) and 
transverse ones (in red) based on the present result in Fig.~\ref{fig:comp} for $\epsilon$ = 1/20, 0.1, 0.5 and 2. Even at the lowest value of $\epsilon$ (1/20) and solid fraction (0.001), the reduction in $\overline{\mathrm{Nu}}$ 
for transverse ridges relative to parallel ones (due to less hydrodynamic slip) is very small. They 
asymptotic solutions for parallel and transverse ridges converge as $\epsilon$ exceeds 1 (because 
$\hat{D}(\epsilon \lambda)$ and $\hat{D}\left(\epsilon \lambda/2\right)$ asymptote to 1/3).
\begin{figure}
\begin{centering}
\includegraphics[width=10cm]{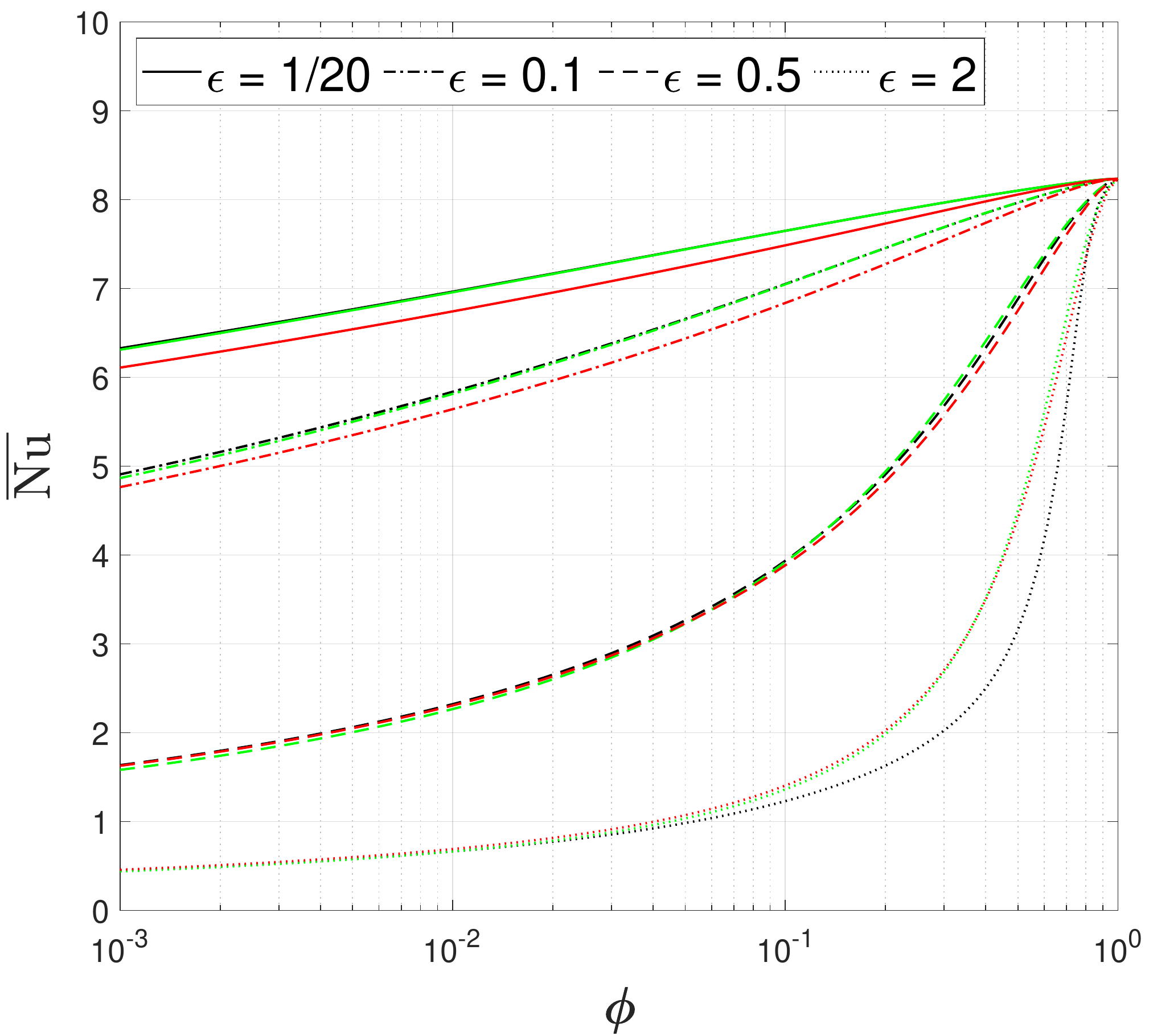}
\par\end{centering}
\protect\caption{$\overline{\mathrm{Nu}}$ versus $\phi$ for parallel ridges (in green based on our asymptotic result,
 \eqref{eq:nupp} and in black based on the exact result by \citet{Kirk-17}) and 
transverse ones (in red) based on the present result for $\epsilon$ = 1/20, 0.1, 0.5 and 2.}
\label{fig:comp}
\end{figure}

\citet{Maynes-13} analytically considered the Graetz-Nusselt form of the 
foregoing problem, focusing on the (periodically) fully-developed limit, as we do here. They utilized a Navier-slip velocity profile in the form of a curve fit to a numerical solution of the actual velocity 
profile (for $\epsilon <$ 2) by \citet{Woolford-09}, which captures interial effects, and thus depends upon the Reynolds number of the flow. This expression was substituted into the
thermal energy equation, which did not include an axial conduction term. Then, the temperature profile was resolved in a form containing an infinite series for a constant heat flux along the ridges. This was utilized to compute the local Nusselt number (also in a form containing an infinite series) as a function of the distance from the inlet to the domain, i.e., where there is a step change in heat flux from zero to a constant value. Eigenvalues and coefficients in the infinite series were evaluated numerically as a function of the normalized slip velocity, i.e., the ratio of the Navier-slip velocity along the composite interface to the mean velocity of the flow, which may be converted to a dimensionless slip length. The first ten of each were tabulated and formulas with constant parameters determined from a least-squares fit approach  provide for equations to calculate an arbitrarily-large number of them. Essentially, this portion of the \citet{Maynes-13} analysis extends the classic Graetz-Nusselt problem resolved by \citet{Cess-59} 
to one which exhibits hydrodynamic, but not thermal, slip.  Duhamel's integral was then used to find the Nusselt number for the problem at hand, where the periodically-varying heat flux is constant over the ridges and zero along menisci. 

We proceed to contrast the approach by \citet{Maynes-13} to our own in order to 
develop the error term which accompanies their mean Nusselt number (denoted by 
$\mathrm{Nu}_{\mathrm{M}}$) in the Stokes flow limit. Insofar as the (periodically fully developed) hydrodynamic problem, \citet{Maynes-13} use the outer solution for the streamwise velocity as per  \eqref{eq:tov} with $G(\epsilon)$ given by  \eqref{eq:geps}, except that 
the \citet{Woolford-09} result alters it as the Reynolds number increases.
Turning to the form of the thermal energy equation used by \citet{Maynes-13}, the axial
conduction term, i.e., the last term on the right side of  \eqref{eq:tebt}, is neglected (c.f.~an extended Graetz-Nusselt problem). Insofar as the outer problem, resolved 
in Section~\ref{sec:ortr}, it does not change the $O(\epsilon^3\mathrm{Re})$ error term in the final
result for the temperature profile,  \eqref{eq:otptr}. Turning to the inner probem treated 
in Section~\ref{sec:itt}, we both ignore the first term on the left side of  \eqref{eq:trte}, introducing an $O(\epsilon^3 \mathrm{Pe})$ error term in  \eqref{eq:oi}. However, we proceed to neglect the 
second and third terms on the left side of  \eqref{eq:trte} and thus introduce an $O(\epsilon^3)$ error term in  \eqref{eq:oi}, whereas \citet{Maynes-13} retain these terms, but ignore the 
axial conduction one and thus introduce an $O(\epsilon)$ error term. By implication, using the \citet{Maynes-13} approach, the error term in the inner temperature profile prior to 
resolving $H(\epsilon)$ is $O(\epsilon,\epsilon^3\mathrm{Re})$. This shows the importance of capturing axial conduction in the inner region.

\section{Conclusions and Recommendations}
We considered laminar, fully-developed, Poiseuille flows of liquid in the Cassie state through diabatic, parallel-plate microchannels symmetrically textured with parallel or transverse isoflux ridges using
matched asymptotic expansions. Our small parameter was ridge pitch divided by microchannel
height. Our slip length result for parallel ridges,  \eqref{eq:bpar}, is well known, but we formally developed it using matched asymptotics as distinct from previous approaches. In the case of the 
standardly-defined mean Nusselt number, we quantified the error in existing expressions and 
provided a new one, \eqref{eq:nupf} which supersedes those in the literature because the error 
term is $O\left(\epsilon^3\right)$ rather than $O\left(\epsilon^2\right)$ and it does not breakdown in
the important limit as solid fraction tends to zero.

We showed that our results for parallel ridges may be directly transformed to those for transverse ones by changing the slip length from $\epsilon \lambda$ to $\epsilon \lambda/2$. However, the error term in the slip length increases from an exponentially small one in $\epsilon$ to $O\left(\epsilon^3 \mathrm{Re}\right)$. Moreover, the error term in the Nusselt number, in addition to an order $O\left(\epsilon^3\right)$ term, has terms that also
depend on Re and Pr as per the error term in \eqref{eq:nuloct}.  Whereas, under the assumptions invoked, the results for parallel ridges are valid 
for any (stable) laminar flow for sufficiently small $\epsilon$, this is not the case for transverse ones
as per their Re and Pe dependence. 

We have neglected thermocapillary stresses along menisci, but, as per the study by \citet{Hodes-17}, they can be substantial in the limiting case of a flat meniscus.  Moreover, \citet{Kirk-20} have shown that to properly resolve them,
meniscus curvature needs to be simultaneously considered. Also, in the case of water, evaporation near the triple contact line and condensation elsewhere along the meniscus will enhance heat 
transfer \citep{Hodes-15}, an 
effect not captured by \citet{Lam-15} in their predicting that water-based microchannel cooling is
degraded by flowing the liquid in the Cassie state. Indeed, thermal energy will conduct from 
the solid-liquid interface into the portion of it near the triple contact line to drive evaporation and then condense elsewhere
along the meniscus. Phase change effects also invalidate the impermeability condition along the meniscus. Until such secondary effects are resolved, it is not clear whether or not water-based 
microchannel cooling may be enhanced by using superhydrophobic surfaces. It's fairly clear that 
microchannel cooling using a liquid metal, such as Galinstan, may be enhanced by using 
superhydrophobic surfaces \citep{Lam-15}. (Phase change effects are not relevant for Galinstan due to its negligible vapor pressure, although it more easily transitions to turbulence on account of its relatively
high density.) Our results enable more accurate prediction of such enhancement. 

\backsection[Acknowledgements]{We acknowledge productive discussions with Dr. Georgios Karamanis on defining the various Nusselt numbers.}

\backsection[Declaration of interests]{The authors report no conflict of interest.}

\bibliographystyle{jfm}
\bibliography{bibliography_v3}

\end{document}